\pdfoutput=1

\documentclass[11pt,twoside,a4paper,cmspaper,final,collab]{cms-tdr}
\def\svnVersion{467043P}\def\cmsCernNoTag{CERN-EP-2017-312}\def\cmsCernDate{\today}\def\cmsMessage{Published in Physics Letters B as \href{http://dx.doi.org/10.1016/j.physletb.2018.06.028}{\doi{10.1016/j.physletb.2018.06.028.}}}

\begin{document}\cmsNoteHeader{SUS-16-040}

\hyphenation{had-ron-i-za-tion}
\hyphenation{cal-or-i-me-ter}
\hyphenation{de-vices}
\RCS$HeadURL: svn+ssh://svn.cern.ch/reps/tdr2/papers/SUS-16-040/trunk/SUS-16-040.tex $
\RCS$Id: SUS-16-040.tex 465751 2018-06-22 17:52:59Z jaehyeok $

\newlength\cmsFigWidth
\ifthenelse{\boolean{cms@external}}{\setlength\cmsFigWidth{0.98\columnwidth}}{\setlength\cmsFigWidth{0.48\textwidth}}
\newlength\cmsFigWidthLarge
\ifthenelse{\boolean{cms@external}}{\setlength\cmsFigWidthLarge{0.60\textwidth}}{\setlength\cmsFigWidthLarge{0.70\textwidth}}
\ifthenelse{\boolean{cms@external}}{\providecommand{\cmsLeft}{top\xspace}}{\providecommand{\cmsLeft}{left\xspace}}
\ifthenelse{\boolean{cms@external}}{\providecommand{\cmsRight}{bottom\xspace}}{\providecommand{\cmsRight}{right\xspace}}

\newcommand\MJ{\ensuremath{M_{\text{J}}}\xspace}
\newcommand\mll{\ensuremath{m_{\ell\ell}}\xspace}
\newcommand\Nb{\ensuremath{N_{\PQb}}\xspace}
\newcommand\Wjets{\ensuremath{\PW\text{+jets}}\xspace}
\newcommand\Zjets{\ensuremath{\PZ\text{+jets}}\xspace}
\newcommand\Nlep{\ensuremath{N_{\text{lep}}}\xspace}
\newcommand\Njet{\ensuremath{N_{\text{jet}}}\xspace}
\newcommand\x{\ensuremath{\phantom{0}}}

\cmsNoteHeader{SUS-16-040}

\title{Search for \texorpdfstring{$R$-parity}{R-parity} violating supersymmetry in pp collisions at \texorpdfstring{$\sqrt{s}=13\TeV$}{sqrt(s) = 13 TeV} using \cPqb~jets in a final state with a single lepton, many jets, and high sum of large-radius jet masses}

\date{\today}

\abstract{
Results are reported from a search for physics beyond the standard model in proton-proton collisions at a center-of-mass energy of $\sqrt{s}=13\TeV$. The search uses a signature of a single lepton, large jet and bottom quark jet multiplicities, and high sum of large-radius jet masses, without any requirement on the missing transverse momentum in an event. The data sample corresponds to an integrated luminosity of 35.9\fbinv recorded by the CMS experiment at the LHC. No significant excess beyond the prediction from standard model processes is observed. The results are interpreted in terms of upper limits on the production cross section for $R$-parity violating supersymmetric extensions of the standard model using a benchmark model of gluino pair production, in which each gluino decays promptly via $ \PSg \to \PQt \PQb \PQs $. Gluinos with a mass below 1610\GeV are excluded at 95\% confidence level.
}

\hypersetup{%
pdfauthor={CMS Collaboration},%
pdftitle={Search for R-parity violating supersymmetry in pp collisions at sqrt(s) =13 TeV using b jets in the single-lepton final state with high jet multiplicity and sum of large-radius jet masses},%
pdfsubject={CMS},%
pdfkeywords={CMS, physics, supersymmetry}}

\maketitle

\section{Introduction}
\label{sec:Introduction}

Searches for physics beyond the standard model (SM) are motivated by several considerations, including theoretical problems associated with explaining the observed mass of the Higgs boson in the presence of quantum corrections (the hierarchy problem)~\cite{BARBIERI198863}, and astrophysical evidence for dark matter~\cite{Bertone:2004pz}.
While the SM has been successful in describing a vast range of phenomena, its inability to address these theoretical and experimental issues makes it an incomplete description of fundamental particles and their interactions.

Supersymmetry (SUSY), a proposed extension of the SM, provides possible solutions to these problems~\cite{ref:hierarchy1,ref:hierarchy2,Ramond:1971gb,Golfand:1971iw,Neveu:1971rx,Volkov:1972jx,Wess:1973kz,Wess:1974tw,Fayet:1974pd,Nilles:1983ge}.
The hierarchy problem can be addressed by SUSY models with a sufficiently low-mass top squark and gluino, and the lightest supersymmetric particle (LSP), if stable, is a potential dark matter candidate~\cite{BARBIERI198863,Dimopoulos:1995mi,Barbieri:2009ev,Papucci:2011wy,Feng:2013pwa}.
That stability is assured in $R$-parity conserving (RPC) SUSY models,
where the $R$-parity of a particle is defined as $(-1)^{2s+3(B-L)}$ with $s$, $B$, and $L$ denoting the spin, baryon number, and lepton number of the particle, respectively~\cite{FARRAR1978575}.

Recent searches at the CERN LHC have set stringent limits on RPC SUSY production, as mass limits for the models studied are reaching ${\sim} 1\TeV$ for the top squark~\cite{Aaboud:2017aeu,Sirunyan:2017xse,Aaboud:2017ayj} and ${\sim} 2\TeV$~\cite{Aaboud:2017bac,Aaboud:2017vwy,Aaboud:2017hdf,0lepsusy_cms_2016,Sirunyan:2017cwe,1lepsusy_cms_2016} for the gluino.
Due to these limits, there is mounting tension in the ability of RPC SUSY models to explain the hierarchy problem with little fine tuning.
These RPC SUSY searches, however, typically require signatures with significant missing transverse momentum (\ptmiss) resulting from the undetected LSPs, while in $R$-parity violating (RPV) SUSY, the LSP is not stable and decays to SM particles, which removes the large \ptmiss signature.
Though this disfavors the LSP as a dark matter candidate, it allows RPV SUSY models to evade constraints from typical RPC SUSY searches.

Given that there is no fundamental theoretical reason for $R$-parity conservation, RPV SUSY yields an important class of models that can ease the tension between natural solutions to the hierarchy problem and current experimental limits.
In addition, the absence of a \ptmiss requirement can allow RPV SUSY searches to be sensitive to a parameter space of RPC SUSY where only a small amount of \ptmiss is expected, such as in models where the mass splitting between the supersymmetric particles is small.
Therefore, RPV SUSY searches help to complete the coverage of SUSY parameter space.

The additional $R$-parity violating terms in the superpotential are

\begin{equation}
\label{eq:RPVLagrangian}
W = \frac{1}{2}\lambda^{ijk}L_{i}L_{j}\overline{\mathrm{e}}_{k}+\lambda^{'ijk}L_{i}Q_{j}\cPaqd_{k}
+\frac{1}{2}\lambda^{''ijk}\cPaqu_{i}\cPaqd_{j}\cPaqd_{k}+\mu^{'i}L_{i}H_{u}.
\end{equation}

\noindent
Here $L_{i}$, $Q_{j}$, and $H_u$ are SU(2) doublets corresponding to leptons, quarks, and the Higgs boson, respectively.
The fields $\overline{\mathrm{e}}_k$, $\cPaqu_i$, and $\cPaqd_j$ are the charged lepton, up-type quark, and down-type quark SU(2) singlets, while the various $\lambda$ and $\mu$ factors denote the coupling strengths for their corresponding interaction.
Color indices are suppressed and letters $i$, $j$, $k$ denote generation indices.
More details on RPV SUSY can be found in Ref.~\cite{Barbier:2004ez}.

This search is motivated by a particular model of $R$-parity violation, minimal flavor violating (MFV) SUSY~\cite{ref:MFV}, in which the $R$-parity violating couplings arise from the SM Yukawa couplings.
This makes the third generation RPV couplings large and those of the first two generations small, which is consistent with the strong experimental constraints from proton decay searches on baryon and lepton number violation involving the lightest two generations~\cite{Barbier:2004ez}.
The coupling $\lambda''^{ijk}$ must be antisymmetric in the last two indices because of gauge invariance, which requires $\lambda''^{\cPqt\cPqb\cPqb}$ to be 0.
Therefore, the largest allowed MFV coupling is $\lambda''^{\cPqt\cPqb\cPqs}$.
 
\begin{figure}[tbp!]
\centering
\includegraphics[width=\cmsFigWidth]{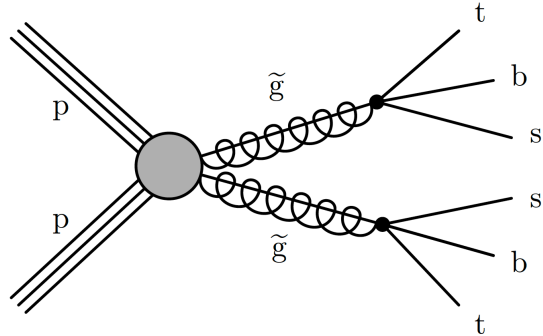}

\caption{Example diagram for the simplified model used as the benchmark signal in this analysis.}
\label{fig:tbs_diagram}
\end{figure}

Due to the high $\PSg\PSg$ cross section and large value of $\lambda''^{\cPqt\cPqb\cPqs}$, a search for the pair production of gluinos that decay via $\PSg \to \cPqt\PASQt\to\cPqt\cPqb\cPqs$ is well motivated.
The simplified model~\cite{bib-sms-2,bib-sms-4} that is used in the interpretation makes several assumptions about the SUSY mass spectrum.
It is assumed that squarks other than the top squark are much heavier than the gluino, so they do not affect the gluino decay,
and the branching ratio of $\PSg \to \cPqt\PASQt\to\cPqt\cPqb\cPqs$ is $100\%$.
The top squark is assumed to be virtual in its decay. This results in a three-body decay,
so searches for dijet resonances, \ie, $\PASQt\to\cPqb\cPqs$, are not applicable in this scenario.
It is further assumed that the gluinos decay promptly.
An example diagram for this simplified model is shown in Fig.~\ref{fig:tbs_diagram}.
Although this benchmark is used for interpreting results, the search  is structured to be generically sensitive to high-mass signatures with large jet and bottom quark jet multiplicities and either little or no \ptmiss, which are potential features of other models of physics beyond the SM.
Previous limits on such MFV models were obtained by the ATLAS and CMS Collaborations at $\sqrt{s}=8\TeV$~\cite{Aad:2016kww,Khachatryan:2016iqn,Khachatryan:2016unx} and by the ATLAS Collaboration at $\sqrt{s}=13\TeV$~\cite{1leprpv_atlas_2016}, excluding gluino masses below ${\sim} 1\TeV$ and $1.6\TeV$, respectively.

This analysis searches in a single-lepton (electron or muon) final state for an excess of events with a large number of identified bottom quark (\cPqb-tagged) jets in regions determined as a function of the jet multiplicity and the sum of masses of large-radius jets, \MJ.
Signal events are expected to contribute to this final state through the leptonic decay of one of the top quarks while populating the high jet multiplicity and high \MJ kinematic regions due to the hadronic decay of the second top quark and the additional bottom and strange quark jets.
The four \cPqb~quarks, two from the top quark decays and two from the top squark decays, provide a high \cPqb-tagged jet multiplicity signature.
The quantity \MJ was proposed in phenomenological studies~\cite{Hook:2012fd,Cohen:2012yc,Hedri:2013pvl}
and was first used for RPC SUSY searches by the ATLAS Collaboration in all-hadronic final states~\cite{Aad:2015lea,Aad:2013wta}
and by the CMS Collaboration in single-lepton events~\cite{Khachatryan:2016uwr,1lepsusy_cms_2016}.

\section{The CMS detector, samples, and event selection}
\label{sec:Samples}

This search uses a sample of proton-proton collision data at a center-of-mass energy of $\sqrt s=13\TeV$ corresponding to an integrated luminosity of 35.9\fbinv, which was collected by the CMS experiment during 2016.
The central feature of the CMS detector is a superconducting solenoid of 6\unit{m} internal diameter,
providing a magnetic field of 3.8\unit{T}. Within the solenoid volume are the charged particle tracking
systems, composed of silicon-pixel and silicon-strip detectors, and the calorimeter systems,
consisting of a lead tungstate crystal electromagnetic calorimeter (ECAL), and a brass and
scintillator hadron calorimeter. Muons are identified and measured by gas-ionization
detectors embedded in the magnetic flux-return yoke outside the solenoid.
A more detailed description of the CMS detector, together with a definition of the coordinate system used and the relevant kinematic variables, is given in Ref.~\cite{Chatrchyan:2008zzk}.

{\tolerance=700
The background predictions use Monte Carlo (MC) simulation samples with corrections to the normalization and shape of distributions measured in data control samples.
\MGvATNLO 2.2.2 is used in leading-order mode~\cite{Alwall:2014hca,Alwall:2007fs} to generate the
\ttbar, \Wjets, quantum chromodynamics multijet (QCD), and Drell--Yan background processes with extra partons.
Comparison to a \POWHEG 2.0~\cite{Nason:2004rx,Frixione:2007vw,Alioli:2010xd} sample generated at next-to-leading order (NLO) shows
that the NLO effects do not have a significant impact.
The $\ttbar$W, $\ttbar$Z, $\ttbar \ttbar$, and $t$-channel single top quark production backgrounds
are generated with \MGvATNLO 2.2.2 in NLO mode~\cite{Frederix:2012ps}, while
the $\cPqt\PW$, $\cPaqt\PW$, and $s$-channel single top quark processes are generated with \POWHEG 2.0.
The \ttbar, \Wjets, and QCD samples are generated with up to 2, 4, 2 extra partons, respectively.
All samples are generated using a top quark mass of 172.5 \GeV and with the NNPDF3.0 set of parton distribution functions (PDF)~\cite{Ball:2014uwa}.
For the fragmentation and showering of partons, the generated samples are interfaced with PYTHIA 8.205~\cite{pythia8.2} and use the CUETP8M1 tune to describe the underlying event~\cite{Skands2014}.
All samples use the highest precision cross sections available~\cite{PhysRevLett.110.252004,Gavin:2012sy,Alioli:2009je,Re:2010bp,Frixione:2015zaa,Bevilacqua:2012em,Nagy:2001fj}.
The detector response is simulated with \GEANTfour~\cite{Agostinelli:2002hh}.
Simulated samples are processed through the same reconstruction algorithms as the data.
\par}

The signal samples are generated with up to two extra partons in leading-order mode and dynamic factorization and renormalization scales by \MGvATNLO~2.2.2.
The same fragmentation, parton showering, simulation, and event reconstruction procedure as for the background samples is used.
 The samples are normalized to NLO + next-to-leading logarithmic cross sections~\cite{XSecgluinogluino}.

The reconstruction of objects in an event proceeds from the candidate particles identified by the particle-flow (PF) algorithm~\cite{CMS-PRF-14-001}, which uses information from the tracker, calorimeters, and muon systems to identify the candidates as charged or neutral hadrons, photons, electrons, or muons.
Charged-particle tracks are required to originate from the event primary vertex (PV), which is the reconstructed vertex with the largest value of summed physics-object squared transverse momentum ($\pt$).
The physics objects used for the PV reconstruction are those returned by a jet finding algorithm~\cite{Cacciari:2008gp,Cacciari:2011ma} with the tracks assigned to the vertex as inputs, and the associated missing transverse momentum, taken as the negative vector sum of the \pt of those objects.

Electrons are reconstructed by pairing a charged-particle track with an ECAL supercluster~\cite{Khachatryan:2015hwa}.
The resulting electron candidates are required to have $\pt>20\GeV$ and $|\eta|<2.5$, and to satisfy identification criteria designed to remove hadrons misidentified as electrons, photon conversions, and electrons from heavy-flavor hadron decays.
Muons are reconstructed by associating tracks in the muon system with those found in the silicon tracker~\cite{Chatrchyan:2012xi}. Muon candidates are required to satisfy $\pt>20\GeV$, $|\eta|<2.4$, and identification criteria designed to select a high-purity muon sample.

To preferentially select leptons that originate in the decay of $\PW$ and $\PZ$ bosons, leptons are required to be isolated from other PF candidates.
The relative isolation of a particle $I^{\rm rel}$ is quantified using an optimized version of the mini-isolation variable $I_{\rm mini}$. Mini-isolation is computed as the scalar sum of the \pt of charged hadrons from the PV, neutral hadrons, and photons that are within a cone of radius $R^{\text{mini-iso}}$ surrounding the lepton momentum vector $\vec{p}^{\, \ell}$ in $\eta$-$\phi$ space~\cite{Rehermann:2010vq}.
The cone radius $R^{\text{mini-iso}}$ varies with $1/\pt^{\, \ell}$ according to
\begin{align}
R^{\text{mini-iso}} &=\begin{cases} 0.2, &
\pt^{\, \ell}\leq 50\GeV\\
{10\GeV}/{\pt^{\, \ell}}, & 50 < \pt^{\, \ell} \leq 200\GeV\\
0.05, & \pt^{\, \ell} > 200\GeV.
\end{cases}\label{eqn:mini_iso_r}
\end{align}
The \pt-dependent cone size reduces the rate of accidental overlaps between the lepton and jets in high-multiplicity or highly Lorentz-boosted events, particularly overlaps between bottom quark jets and leptons originating from a boosted top quark.
Relative isolation is computed as $I^{\rm rel}=I_{\rm mini}/\pt^{\, \ell}$ after subtraction of the average contribution from additional proton-proton collisions in the same bunch-crossing (pileup).
To be considered isolated, electrons and muons must satisfy $I^{\rm rel}<0.1$ and $0.2$, respectively, where the different thresholds account for purity differences between electrons and muons.

The combined efficiency for the electron reconstruction, identification, and isolation requirements is about 50\% at $\pt^{\, \ell}$ of $20\GeV$, increasing to 65\% at $50\GeV$, and reaching a plateau of 80\% above $200\GeV$.
The corresponding efficiency for muons is about 70\% at $\pt^{\, \ell}$ of $20\GeV$, increasing to 80\% at $50\GeV$, and reaching a plateau of 95\% for $\pt^\ell > 200\GeV$. Data-to-simulation corrections (scale factors) are applied for both electrons and muons to correct the simulated lepton selection efficiency to match that observed in data.

The charged PF candidates associated with the PV and the neutral PF candidates are
clustered into jets using the anti-$\kt$ algorithm~\cite{Cacciari:2008gp}
with distance parameter $R=0.4$, as implemented in the \textsc{fastjet} package~\cite{Cacciari:2011ma}.
The estimated contribution to the jet $\pt$
from neutral PF candidates produced by pileup
is removed with a correction based on the area
of the jet and the average energy density of the event~\cite{Cacciari:2007fd}.
The jet energy is calibrated using $\pt$- and $\eta$-dependent corrections; the resulting
calibrated jets are selected if they satisfy $\pt>30\GeV$ and $|\eta|\leq2.4$.  Each
jet must also meet loose identification requirements~\cite{Chatrchyan:2011ds}
to suppress, for example, calorimeter noise. Finally, jets that have PF
constituents matched to the selected lepton are removed from
the jet collection.
These resulting jets are considered to be ``small-$R$'' jets.

The combined secondary vertex algorithm~\cite{Chatrchyan:2012jua,Sirunyan:2017ezt} is applied to each small-$R$ jet to create a subset of \cPqb-tagged jets.
The tagging efficiency for \cPqb~jets in the range $\pt=30$ to $50\GeV$
is 60--67\% (51--57\%) in the barrel (endcap) and increases with \pt.
Above $\pt\approx 150\GeV$ the efficiency decreases to ${\approx} 50$\%.
The probability to misidentify jets arising from $\cPqc$ quarks
is 13--15\% (11--13\%) in the barrel (endcap), while the misidentification probability for light-flavor quarks
or gluons is 1--2\%.
Data-derived scale factors for the \cPqb~tag efficiency and mistag rate are applied to simulation such that the simulated \cPqb~tagging performance matches that observed in data.

\begin{figure}[tbp!]
\centering
\includegraphics[width=\cmsFigWidth]{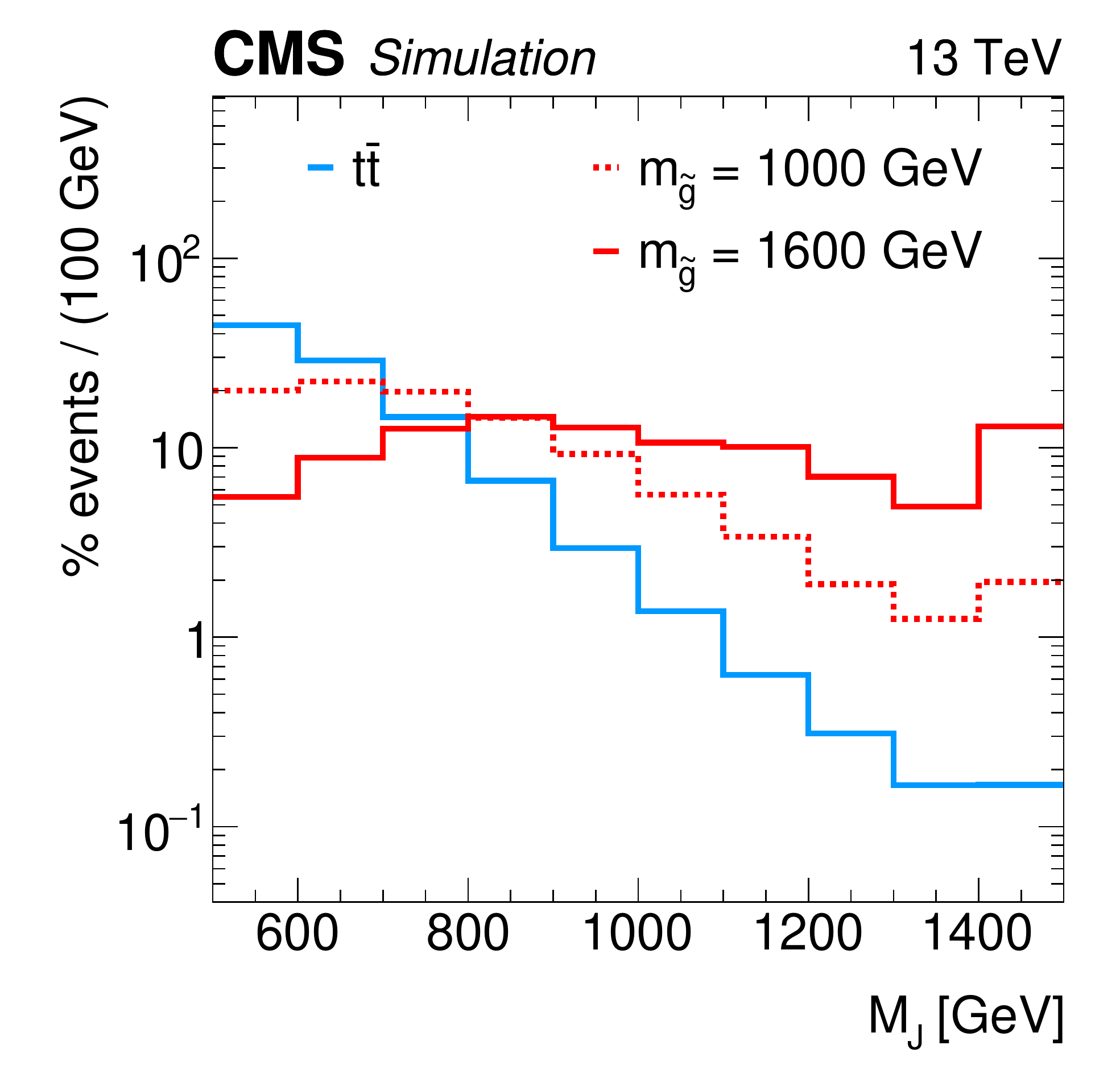}

\caption{Distributions of \MJ, normalized to the same area, for \ttbar events and signal events with two different gluino masses in a selection of $\HT>1200\GeV$, $\Nlep=1$, $\Njet\geq8$, $\MJ>500\GeV$, and $\Nb\geq1$.}
\label{fig:mj_shape}
\end{figure}

``Large-$R$'' ($R=1.2$) jets are created by clustering small-$R$ jets and the selected lepton using the anti-$\kt$ algorithm.
Leptons are included to encompass the full kinematics of the event.
Clustering small-$R$ jets instead of PF candidates incorporates
the jet pileup corrections, thereby reducing the dependence of the large-$R$ jet mass on pileup.
This technique of clustering small-$R$ jets into large-$R$ jets has been used
previously, e.g. Refs.~\cite{Aad:2014bva,Khachatryan:2016uwr,Aaboud:2017aeu}.
The variable \MJ is defined as the sum of all large-$R$ jet masses, where $m(J)$ is the mass of a single large-$R$ jet:
\begin{eqnarray}
\MJ\,\,= \!\!\sum_{J_i\, \in\, {\text{large-}R\text{ jets}}} m(J_i).
\end{eqnarray}
The quantity \MJ is used as a measure of the mass-scale of an event.
Signal events tend to have large \MJ as the large-$R$ jets capture the kinematic information of the high-mass gluinos.
Comparatively, SM background processes tend to have smaller values of \MJ due to their lower mass-scales.
SM events, however, can have large values of \MJ in the presence of significant initial-state-radiation (ISR).
For example, in \ttbar events, ISR jets can either overlap with \ttbar daughter jets or boost the \ttbar system such that the system is collimated, both of which result in high-mass large-$R$ jets and, correspondingly, high \MJ.
The \MJ distributions for \ttbar and signal are shown in Fig.~\ref{fig:mj_shape}, which uses events with $\Njet \geq 8$ to ensure similar \Njet distributions for both \ttbar and signal.

Events are selected with triggers~\cite{Khachatryan:2016bia} that require either at least one jet with $\pt > 450\GeV$ or the scalar sum of the \pt of all small-$R$ jets (\HT) above $900\GeV$. Trigger efficiencies are over $99\%$ for signal events passing the analysis selection defined below.

These events are further selected with a baseline requirement of exactly one electron or muon, $\MJ>500\GeV$, $\HT>1200\GeV$, that the number of small-$R$ jets (\Njet) be at least 4, and that the number of those jets that are tagged as bottom quark jets (\Nb) be at least~1.

\section{Background prediction}
\label{sec:Overview}

After the baseline selection, the dominant background contribution is from the $\ttbar$ process, with small contributions from \Wjets~and QCD events with a misidentified lepton.
Rare background contributions, classified below as ``Other'', come from single top quark, $\ttbar\PW$, $\ttbar\cPZ$, $\ttbar\PH$, $\ttbar\ttbar$, and Drell--Yan production.

To search for signal events arising from new high-mass particles decaying with large jet and \cPqb-jet multiplicities, the \Nb distribution is examined in different kinematic regions based on \Njet and \MJ.
The \Njet bins are defined to be 4--5, 6--7, and $\geq 8$. The \MJ bins are $500 < \MJ \leq 800$\GeV, $800 < \MJ \leq 1000$\GeV, and $\MJ > 1000$\GeV, with the two highest \MJ~bins merged for the $4 \le \Njet \le 5$ case due to the limited data sample size in the $\MJ > 1000$\GeV region.
The low-\Njet, low-\MJ bins are expected to be background-dominated and are used as control regions to constrain systematic uncertainties, while the high-\Njet, high-\MJ bins are used as signal regions.
A diagram representing this binning is shown in Figure~\ref{fig:analysis_regions}.
The \Nb distribution is separated into $\Nb=1$, 2, 3, and ${\ge}4$ bins for each region.
The two highest \Nb~bins are the most sensitive to signal due to larger signal-to-background ratios, while the lower \Nb~bins provide constraints on the background normalizations and systematic uncertainties.
The signal efficiency for the bin requiring $\Njet \geq 8$ and $\MJ>1000$\GeV is $2\%$ and $8\%$ for $m_{\PSg}= 1000\GeV$ and 1600\GeV, respectively.

A global maximum-likelihood fit is performed to obtain predictions for the SM background processes.
This fit is carried out both for a background-only hypothesis and for signal-plus-background hypotheses, in which an additional signal contribution is extracted.
The model is constructed using the poisson probabilities of the bin contents of the \Nb distribution for all \Njet, \MJ regions, while systematic uncertainties are applied as nuisance parameters.
The \Nb shape for each process is taken from simulation, but varied to assess the impact of mismodeling of relevant parameters, including the rate of gluon splitting to \bbbar and tagging efficiencies for heavy- and light-flavor jets~\cite{Chatrchyan:2012jua,Sirunyan:2017ezt}.
The appropriate ranges for these parameters are determined based on measurements in dedicated control samples and then constrained by a simultaneous fit across all bins of \Njet and \MJ in a correlated manner. Various studies with simulated pseudo-experiments were conducted to validate the likelihood model and to confirm that signal contamination effects are negligible.

Because the kinematic tails of the \Njet and \MJ variables are difficult to model reliably, the \ttbar and QCD normalizations are individually allowed to freely vary in each (\Njet, \MJ) bin.
The \ttbar normalizations are constrained in each bin by the background-dominated $\Nb \leq 2$ bins, while the QCD normalizations are constrained by control regions with no identified leptons ($\Nlep=0$).
These $\Nlep=0$ control regions follow the same kinematic binning as the $\Nlep=1$ bins, but are integrated in \Nb for $\Nb \ge 1$ and use offset \Njet~bins of 6--7, 8--9, and ${\ge}10$ to account for differences in the \Njet distributions between the $\Nlep=1$ and $\Nlep=0$ samples.
The QCD contribution in a particular $\Nlep=1$ bin is then constrained by the corresponding $\Nlep=0$ bin.
To avoid biasing the normalization measurement, the small contribution of \ttbar background to the $\Nlep=0$ control regions is included using the normalization from the corresponding $\Nlep=1$ bins, while contributions from other processes are taken from simulation.

\begin{figure}[tbp!]
\centering
\includegraphics[width=\cmsFigWidth]{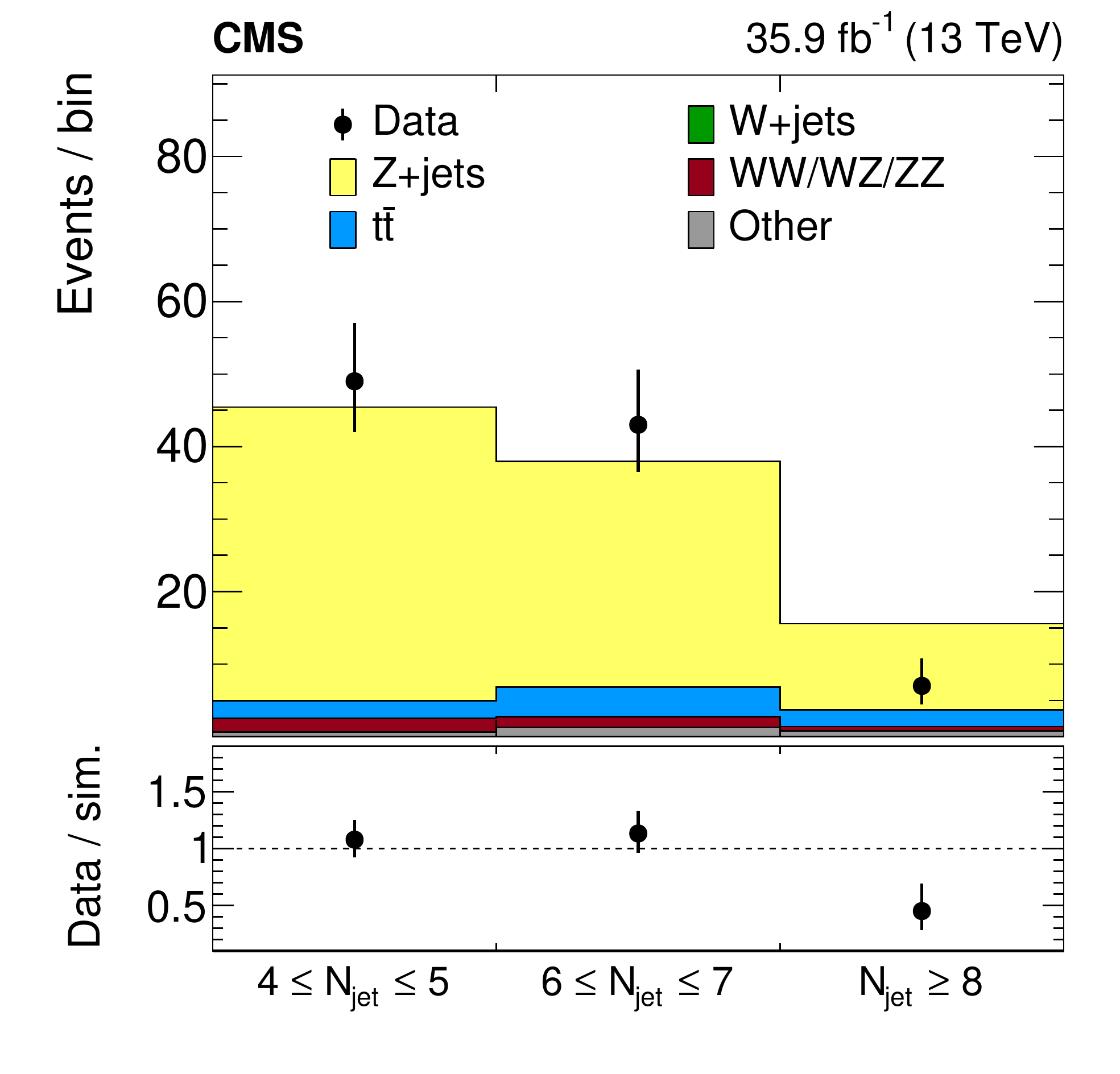}
\caption{Jet multiplicity distribution for data and MC simulation in a \Zjets control sample selected by requiring $\Nlep=2$, $\HT>1200\GeV$, $\MJ>500\GeV$, $\Nb=1$, and $80<\mll<100\GeV$. The total yield from simulation is normalized to the number of events in data.
The uncertainty in the ratio of data to simulation yields (lower panel) is statistical only.}
\label{fig:njets_dy}
\end{figure}

The \Njet shape of the \Wjets background is taken from simulation and allowed to vary based on the data-to-simulation agreement in a kinematically similar \Zjets sample selected with $\Nlep=2$ ($\Pe\Pe$ or $\PGm\PGm$), $\HT>1200\GeV$, $\MJ>500\GeV$, $\Nb=1$, and $80<\mll<100\GeV$, where $\mll$ is the invariant mass of the two leptons.
The \Njet distribution and data/simulation yields ratio for this sample are shown in Fig.~\ref{fig:njets_dy}.
The \Wjets background is then determined in the fit with one global normalization parameter and two parameters to adjust the bin-to-bin normalization based on the difference between the ratios in adjacent $\Njet$ bins -- 17\% between $4\le\Njet\le5$ and $6\le\Njet\le7$ and 62\% between $6\le\Njet\le7$ and $\Njet\ge8$.
After correcting the \Njet spectrum, the residual \MJ mismodeling is expected to be small, so no further correction is applied.

The ``Other'' component is estimated from simulation. Its contribution is less than 20\% of the total backgrounds in all kinematic regions considered.

\begin{figure}[tbp!]
\centering
  \begin{tabular}{ |c|c|c|c| }
    \hline
    \multirow{2}{*}{$M_\text{J}$ [\GeVns{}]} & \multicolumn{3}{c|}{$N_\text{jet}$} \\ \cline{2-4}
                                         & 4--5               & 6--7 & $\geq$8 \\ \hline
    500--800                            & CR                  & CR    & SR \\ \hline
    800--1000                           & \multirow{2}{*}{CR} & SR    & SR \\ \cline{1-1} \cline{3-4}
    $>$1000                              &                     & SR    & SR \\ \hline
  \end{tabular}
  \caption{\label{fig:analysis_regions} Illustration depicting the (\Njet, \MJ) binning after the baseline selection, with control and signal region bins denoted by ``CR'' and ``SR'', respectively.}
\end{figure}

\section{Systematic uncertainties}
\label{sec:systematics}

\subsection{Background systematic uncertainties}

The nominal simulated shape of the \Nb distribution is allowed to vary by the inclusion of systematic uncertainties.
Each uncertainty is incorporated in the fit with template \Nb histograms to account for the effects of the systematic variation and a nuisance parameter $\theta$ to control the variation amplitude.
The nuisance parameters are subject to Gaussian constraints, normalized so that $\theta=0$ corresponds to the nominal \Nb shape and $\theta=\pm1$ corresponds to ${\pm}1$ standard deviation (s.d.) variation of the systematic uncertainty.
These uncertainties affect only the \Nb shape for \ttbar, QCD, and W+jets backgrounds, because their normalizations are determined from data, while for the other (subleading) backgrounds the uncertainties affect both the \Nb shape and normalization.

The primary source of systematic uncertainty is the modeling of the gluon splitting rate, which can produce additional \cPqb~quarks in events and may not be properly simulated.
To account for this, a nuisance parameter controlling the gluon splitting rate is included in the likelihood.
The size of the ${\pm}1$ s.d.\ variation for this nuisance parameter is estimated using a fit to the $\Delta R_{\bbbar}$ distribution in a control sample, where $\Delta R_{\bbbar}$ is defined as the $\Delta R$ between two \cPqb-tagged jets in the event.
This control sample is selected by requiring $\Nlep = 0$, $\HT > 1500\GeV$, $\Nb = 2$, $\Njet \geq 4$, and $\MJ >500\GeV$, as the gluon splitting signal in a $\Nlep = 1$ control region is contaminated by \cPqb~quarks from the decay of top quarks.
To ensure that these measurements in the QCD-dominated $\Nlep=0$ region are applicable to the ttbar-dominated $\Nlep=1$ region, both processes are simulated with the same procedure and settings. 
Furthermore, the $\Nlep=0$ control sample is formed from a subset of the data that is selected to be most stable in the \cPqb~tagging algorithm performance, since the precision of the $\Delta R_{\bbbar}$ fit is not limited by the data sample size.
This choice isolates the physical effects of gluon splitting from the potential time dependence of the \cPqb~tagging performance due to variations in experimental conditions, which are separately incorporated by the \cPqb-tag scale factor uncertainties.
The nuisance parameter obtained from this control sample is allowed to vary in the full likelihood fit and further constrained by the observed data in the $\Nlep = 1$ regions.

Events where both of the \cPqb-tagged jets originate from one gluon splitting populate the low-$\Delta R_{\bbbar}$ region, while events without a gluon splitting or where the splitting yields one or no \cPqb-tagged jets populate both the low- and high-$\Delta R_{\bbbar}$ regions roughly equally.
Gluon splittings can sometimes be reconstructed with fewer than two \cPqb-tagged jets either because the quarks are collimated into a single jet, one of the \cPqb~jets is not tagged, or because one of the quarks is not within the kinematic acceptance.

A fit to the $\Delta R_{\bbbar}$ distribution is used to extract the relative contributions of events with and without gluon splitting and is performed in four equal bins in the range $0 \leq \Delta R_{\bbbar} < 4.8$.
This binning is chosen to avoid relying on the fine details of the simulated $\Delta R_{\bbbar}$ shape.
The instances of gluon splitting in simulation are identified by requiring a gluon with $\pt > 30\GeV$ that decays to \cPqb~quarks. Three categories are then defined: events with gluon splitting resulting in two \cPqb-tagged jets (denoted GSbb), with gluon splitting resulting in one or fewer \cPqb-tagged jets (GSb), and without any gluon splitting (no GS).
In the fit, the GSbb and GSb contributions are varied together with a single normalization parameter.

\begin{figure}[tbp!]
\centering
\includegraphics[width=\cmsFigWidth]{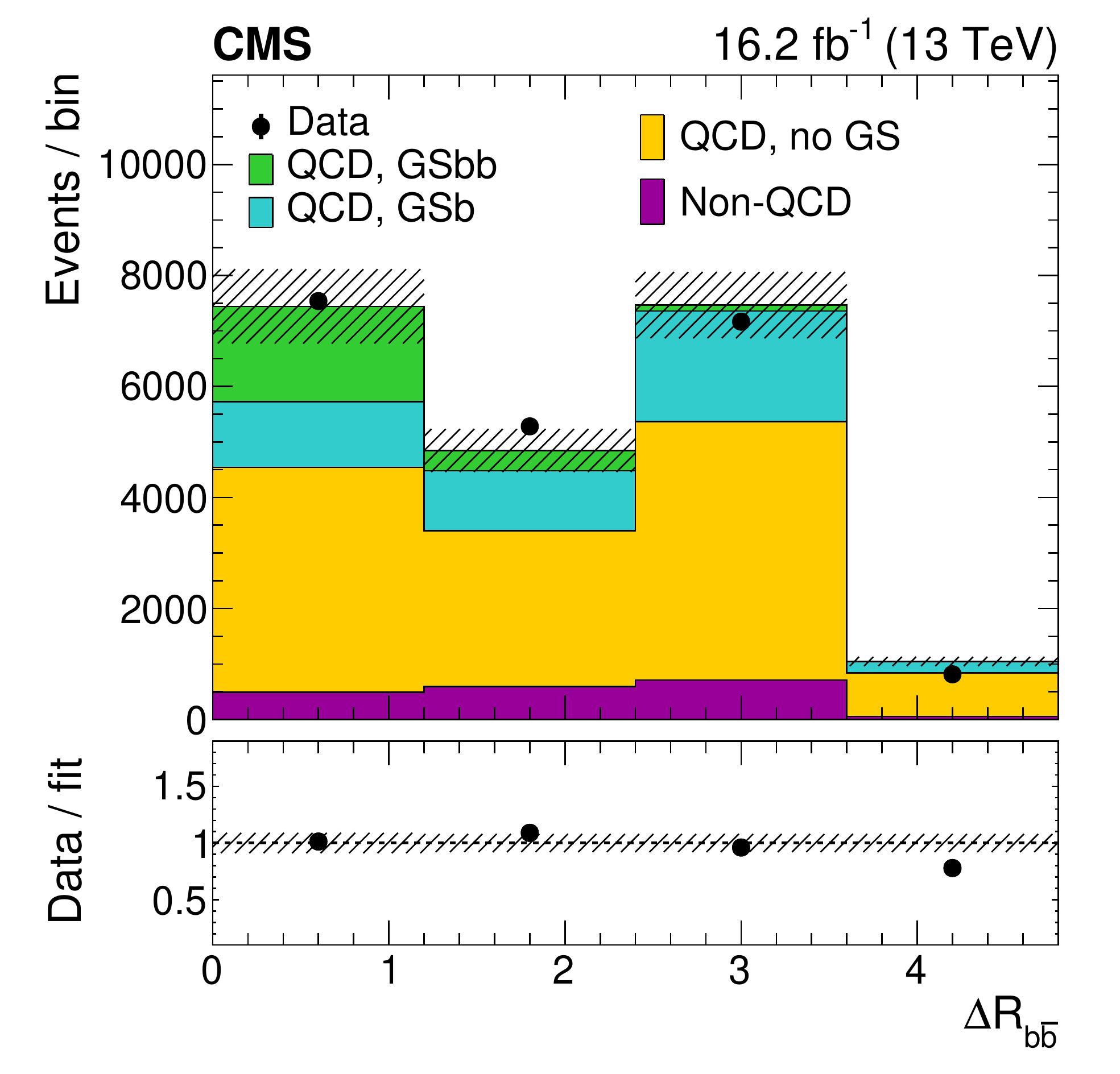}

\caption{Post-fit $\Delta R_{\bbbar}$ distributions in a selection with  $\Nlep = 0$, $\HT > 1500\GeV$, $\Nb = 2$, $\Njet \geq 4$, and $\MJ>500\GeV$ with the post-fit uncertainty represented by a hatched band.
The ratio of data to simulation yields is shown in the lower panel.}
\label{fig:drbb_fitresults}
\end{figure}

The $\Delta R_{\bbbar}$ fit extracts a weight of $0.77 \pm 0.09$ for gluon splitting events and a weight of $1.21 \pm 0.08$ for non-gluon splitting events. The post-fit distributions are shown in Figure~\ref{fig:drbb_fitresults}.
The GSbb and GSb categories are plotted separately to demonstrate the difference in shapes. The discrepancy in the last bin does not significantly impact the fit because the higher yield bins at lower values of $\Delta R_{\bbbar}$ constrain the fit.
The deviations of these weights from unity, summed in quadrature with their post-fit uncertainty, are used to form the ${\pm} 1$~s.d.\ variations of the gluon splitting rate nuisance parameter by applying weights of $1 \pm 0.25$ to gluon splitting events and $1 \mp 0.22$ to non-gluon splitting events in an anti-correlated manner.
The fit results are used as a measure of the uncertainty on modelling of the GS rate as opposed to a correction to the central value, since the $\Delta R_{\bbbar}$ variable may not be a perfect proxy for the GS rate.

Various tests are conducted to assess the stability of the fit results.
To test the dependence of the gluon splitting weights across kinematic regions, the fit is repeated both with a higher \MJ threshold and with different \Njet bins. Additionally the fit is conducted with finer binning to test the dependence of the results on the binning of the $\Delta R_{\bbbar}$ distribution.
The resulting weights are all consistent with those of the nominal fit.

Another significant systematic uncertainty is the uncertainty in the data-to-simulation scale factors (SF) for \cPqb~tagging efficiency and mistag rates.
These scale factors are derived from data in various QCD and \ttbar control samples and are binned in jet \pt and jet flavor (light + \cPg, \cPqc, and \cPqb)~\cite{CMS-DP-2017-012}.
The ${\pm} 1$~s.d. \Nb templates for these scale factors are assessed by varying them according to the uncertainties in their measurements.

Other experimental uncertainties are small and include lepton selection efficiency, lepton misidentification rate, jet energy scale, jet energy resolution, and integrated luminosity.
The uncertainty associated with lepton selection efficiency is determined by varying the efficiency to select a lepton within its uncertainty determined from data.
The $\Nlep$ distribution for QCD events may not be simulated well because it relies on modeling the tail of the fragmentation function and various detector effects. To account for this, an uncertainty of 20\% is assigned to the relative normalization of QCD events in the 0- and 1-lepton bins, which is motivated by data-to-simulation studies of lepton isolation distributions.
Jet energy scale uncertainties~\cite{Chatrchyan:2011ds,1748-0221-12-02-P02014} are assessed by varying the \pt of small-$R$ jets as a function of \pt and $\eta$. The uncertainty arising from jet energy resolution ~\cite{Chatrchyan:2011ds,1748-0221-12-02-P02014} is determined by applying an $\abs{\eta}$-dependent factor to the jet \pt to match the jet energy resolution observed in data.
The integrated luminosity is varied according to its uncertainty of 2.5\%~\cite{CMS-PAS-LUM-17-001}, affecting only the backgrounds estimated from simulation.
No uncertainty is applied for the amount of pileup as studies have shown its effect to be negligible in this high-\HT selection.
The uncertainties due to the limited size of simulation samples are incorporated as uncorrelated nuisance parameters in the fit.

Theoretical systematic uncertainties are applied and include independent and correlated variations of the renormalization  and factorization scales.
Additionally, uncertainties on the PDF are incorporated by considering variations in the NNPDF 3.0 scheme~\cite{Ball:2014uwa}.
The size of these uncertainties is typically small as the effect of these variations is largely to modify the cross section of processes, which for the main backgrounds are constrained by data.

\begin{figure*}[tbp!]
\centering
\includegraphics[width=\cmsFigWidth]{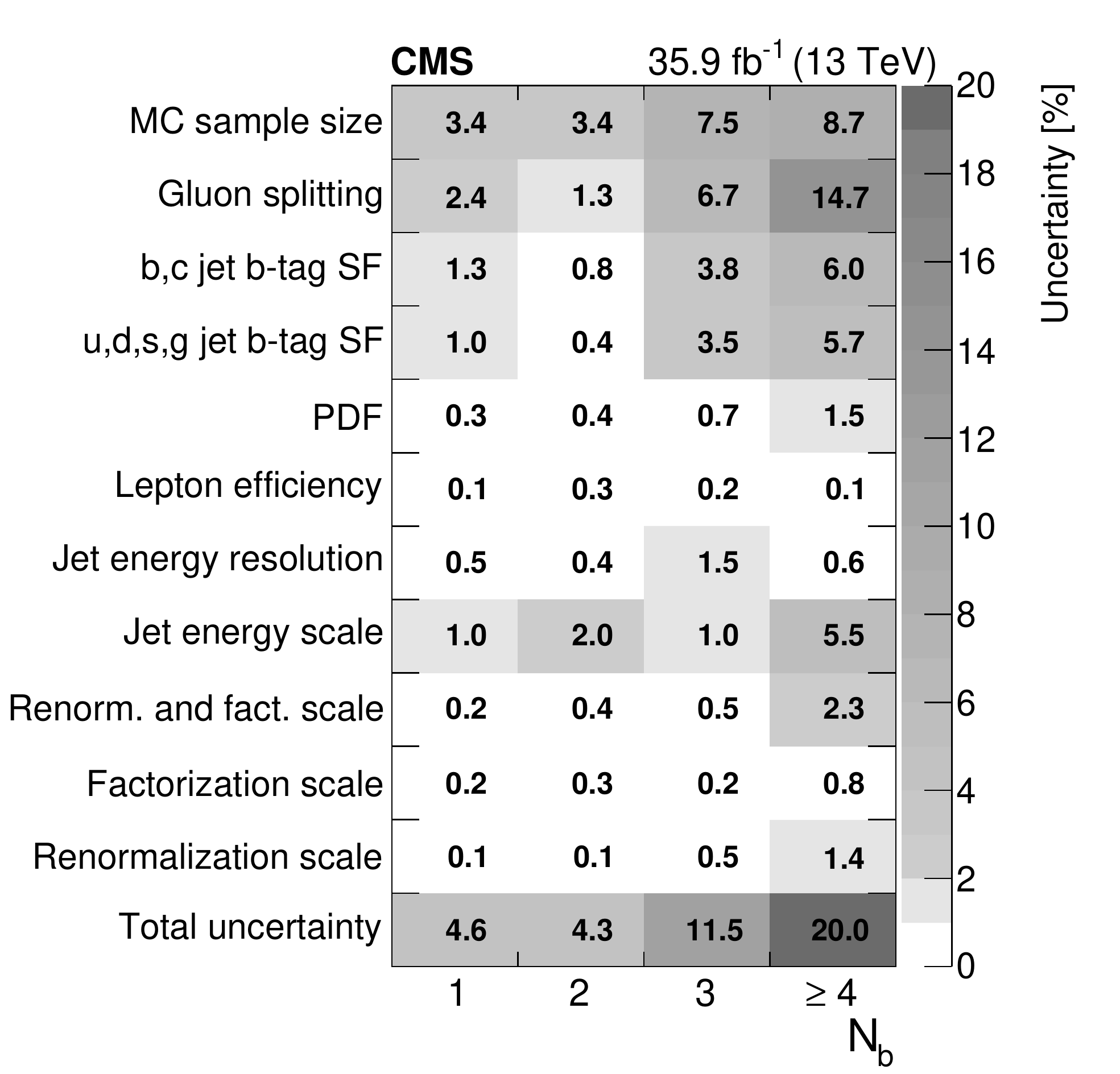}
\includegraphics[width=\cmsFigWidth]{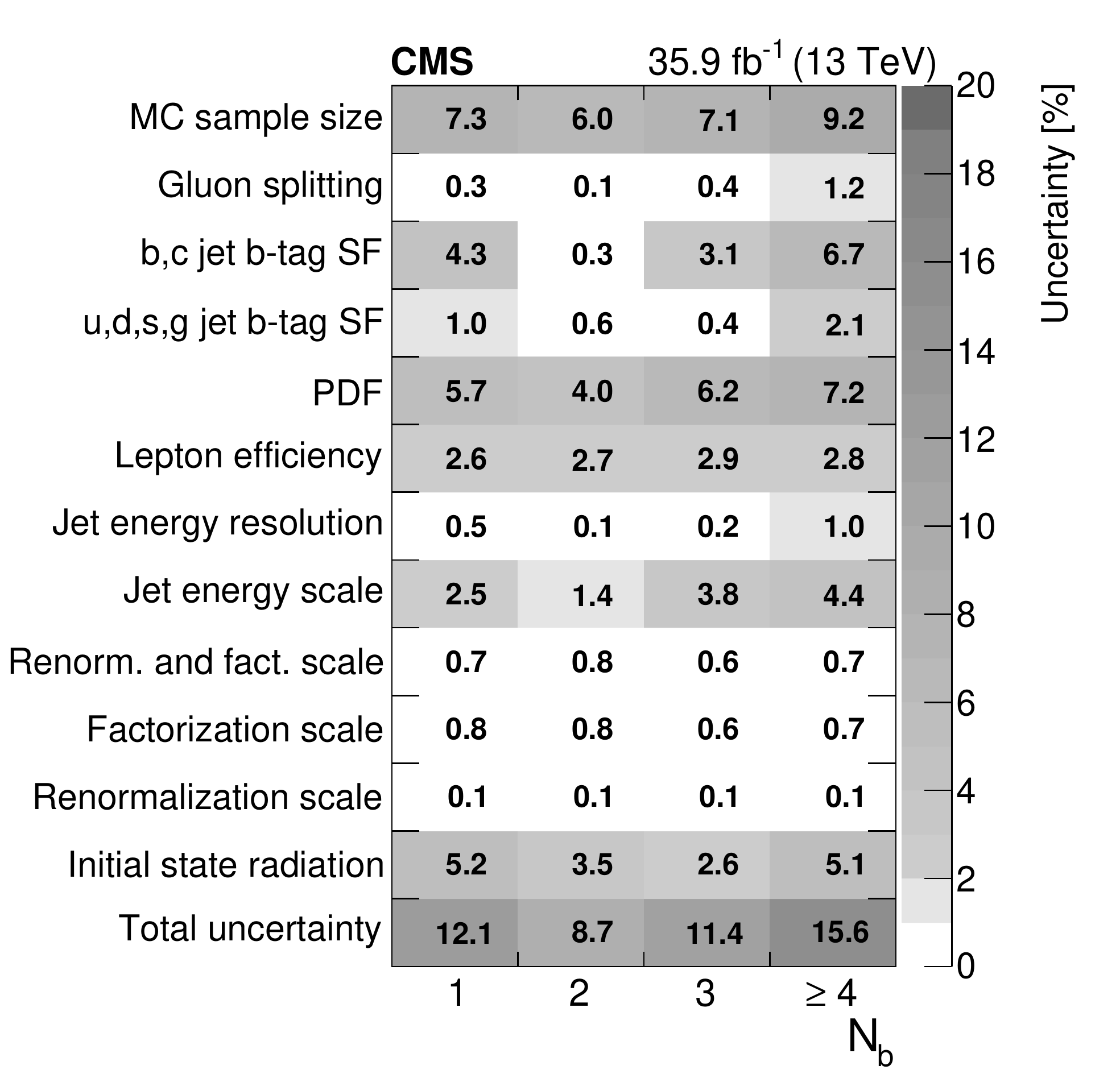}

\caption{Background (left) and $m_{\PSg}=1600\GeV$ signal (right) systematic uncertainties affecting the \Nb shape (in percent) in the $\Njet \geq 8$ and $\MJ \geq 1000\GeV$ bin. The bottom row shows the total uncertainty for a given \Nb bin by summing in quadrature all uncertainties. These values are similar for other (\Njet, \MJ) bins.}
\label{fig:SystematicsNlep1}
\end{figure*}

The background systematic uncertainties that affect the \Nb shape are shown in Fig.~\ref{fig:SystematicsNlep1} (left) for the most sensitive search bin.

\subsection{Signal systematic uncertainties}

Several of the systematic uncertainties affecting the signal yield are evaluated in the same way as the background yield.
These are the uncertainties due to gluon splitting, lepton selection efficiency, jet energy scale, jet energy resolution, \cPqb~tagging scale factors, simulation sample size, integrated luminosity, and theoretical uncertainties.
All systematic variations affect both the \Nb shape and normalization, except for the gluon splitting uncertainty, which is taken to affect only the \Nb shape.

The number of jets from ISR produced in the signal simulation is reweighted based on comparisons between data and simulated \ttbar samples.
The reweighting factors vary between 0.92 and 0.51 for the number of ISR jets between 1 and ${\ge}6$.
One half of the deviation from unity is taken as the systematic uncertainty in these reweighting factors.

The systematic uncertainties affecting the signal \Nb shape are shown in Fig.~\ref{fig:SystematicsNlep1} (right) for the most sensitive bin in a model with $m_{\PSg}=1600\GeV$.
The dominant signal systematic uncertainties arise from the limited simulation sample size, the \cPqb~tagging efficiency scale factors, and the ISR modeling.
There is no systematic uncertainty taken for pileup reweighting, as the signal efficiency is found to be insensitive to the number of pileup interactions.

\section{Results}
\label{sec:Results}

\begin{figure*}[tbp!]
\centering
\includegraphics[width=\cmsFigWidth]{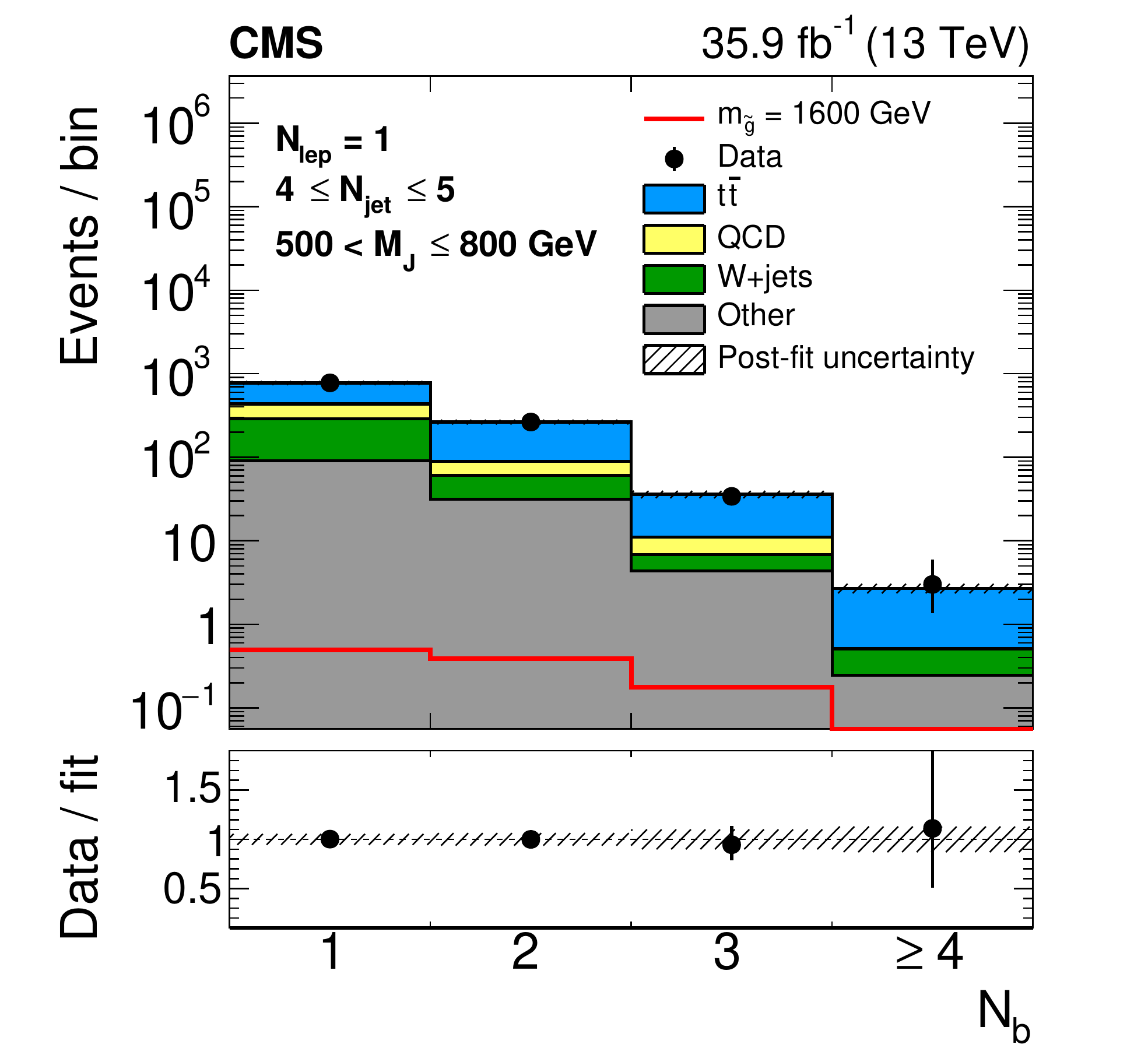} \hfil
\includegraphics[width=\cmsFigWidth]{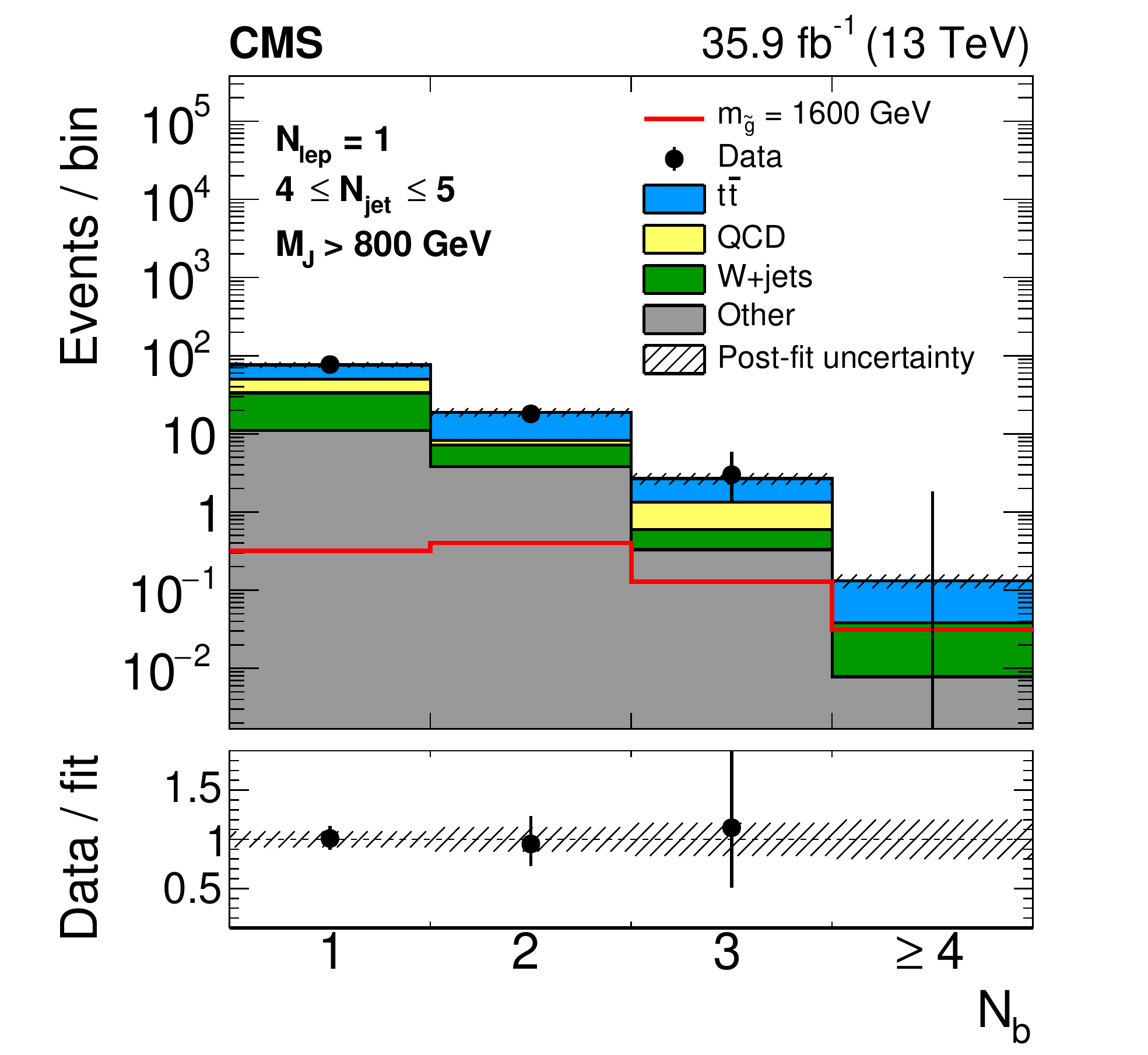} \\
\includegraphics[width=\cmsFigWidth]{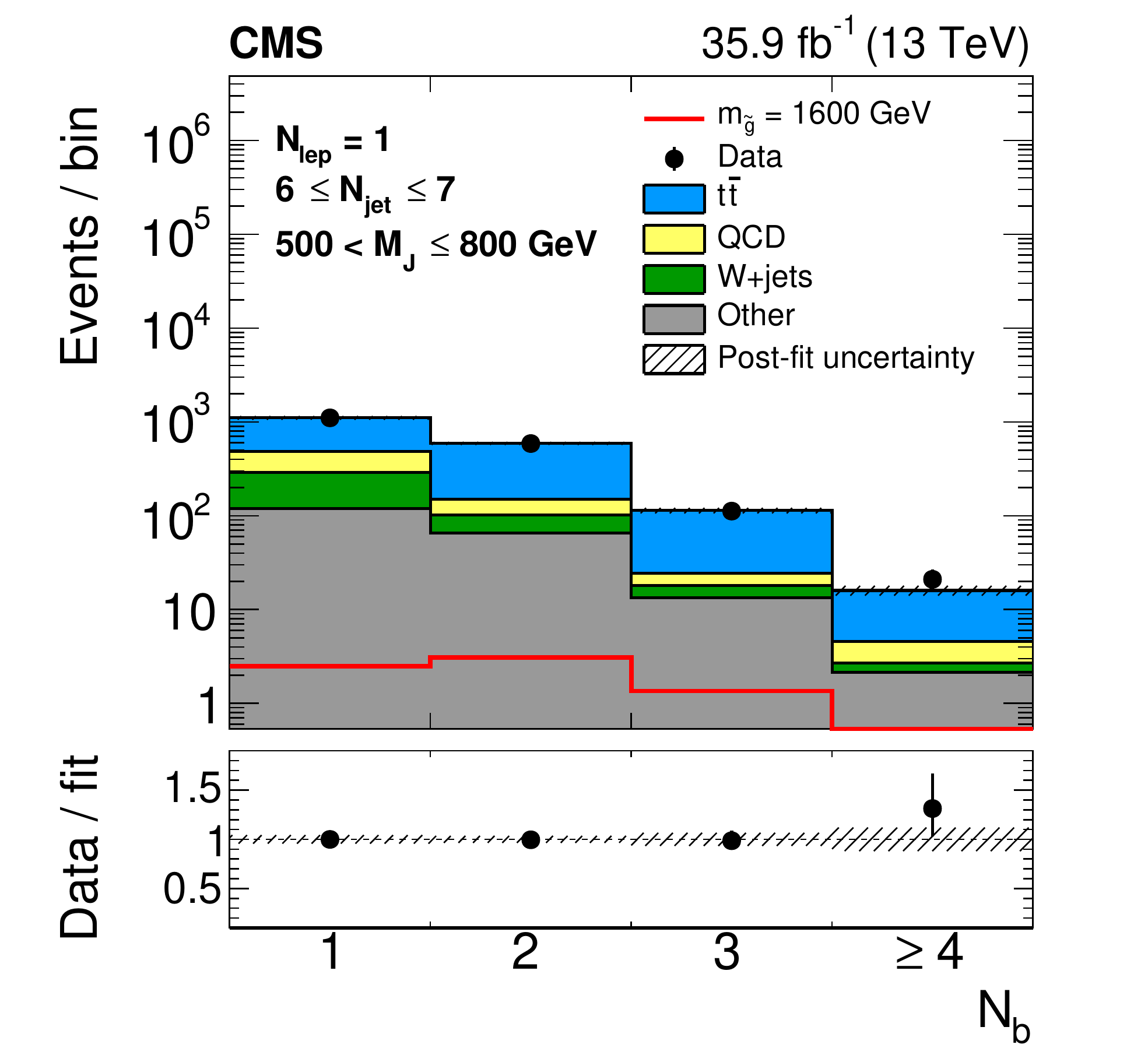} \hfil
\includegraphics[width=\cmsFigWidth]{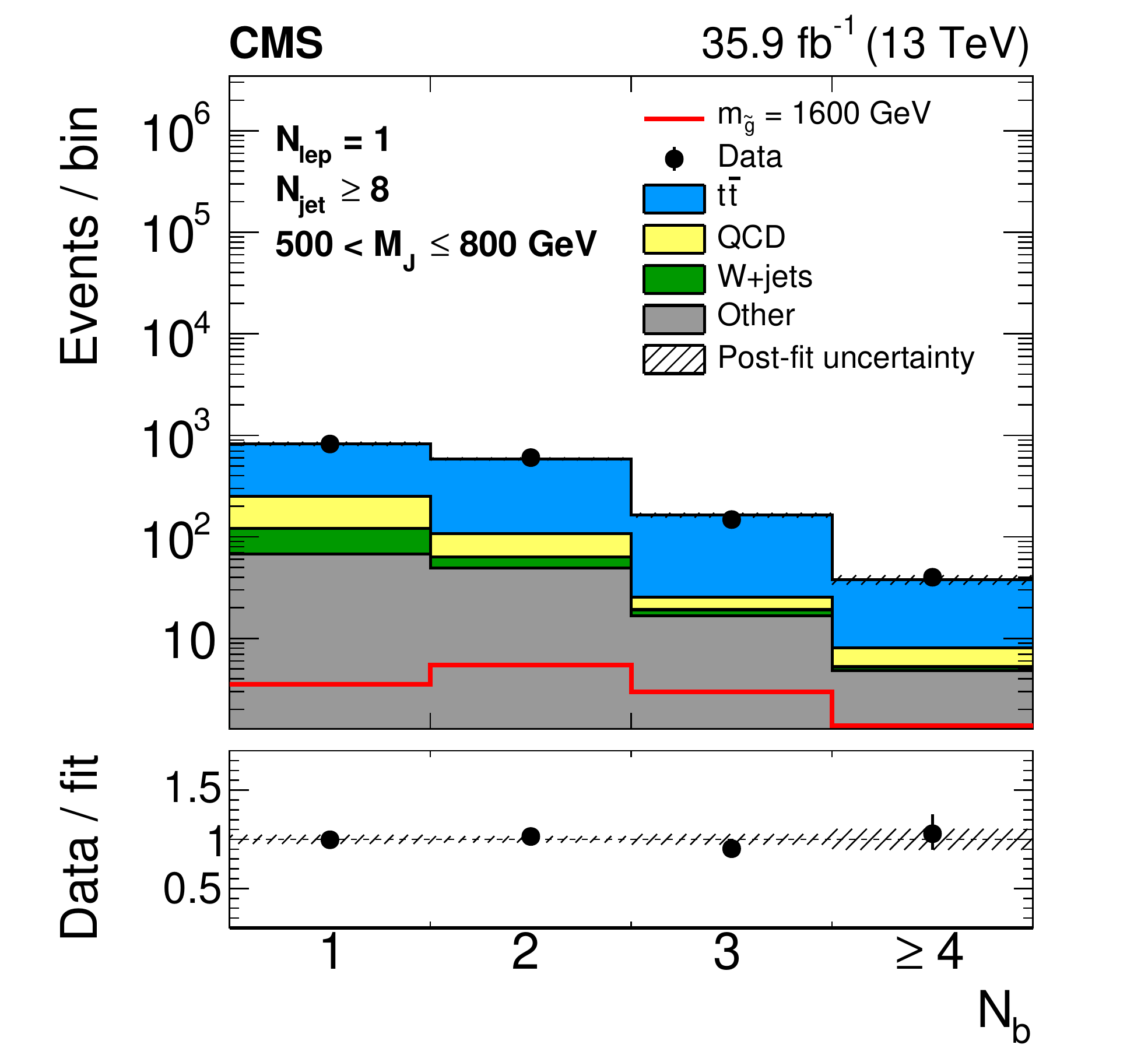}

\caption{Data and the background-only post-fit \Nb distribution for bins with low expected signal contribution: $500 < \MJ \leq 800\GeV$, $ 4 \leq \Njet \leq 5$ (upper-left), $\MJ > 800\GeV$, $4 \leq \Njet \leq 5$ (upper-right), $500 < \MJ \leq 800\GeV$, $6 \leq \Njet \leq 7$ (lower-left), and $500 < \MJ \leq 800\GeV$, $\Njet \geq 8$ (lower-right).
The expected signal distribution is also shown for a gluino mass of 1600\GeV.
The ratio of data to post-fit yields is shown in the lower panel.
The post-fit uncertainty is depicted as a hatched band.}
\label{fig:bonly_cr}
\end{figure*}

\begin{figure*}[tbp!]
\centering
\includegraphics[width=\cmsFigWidth]{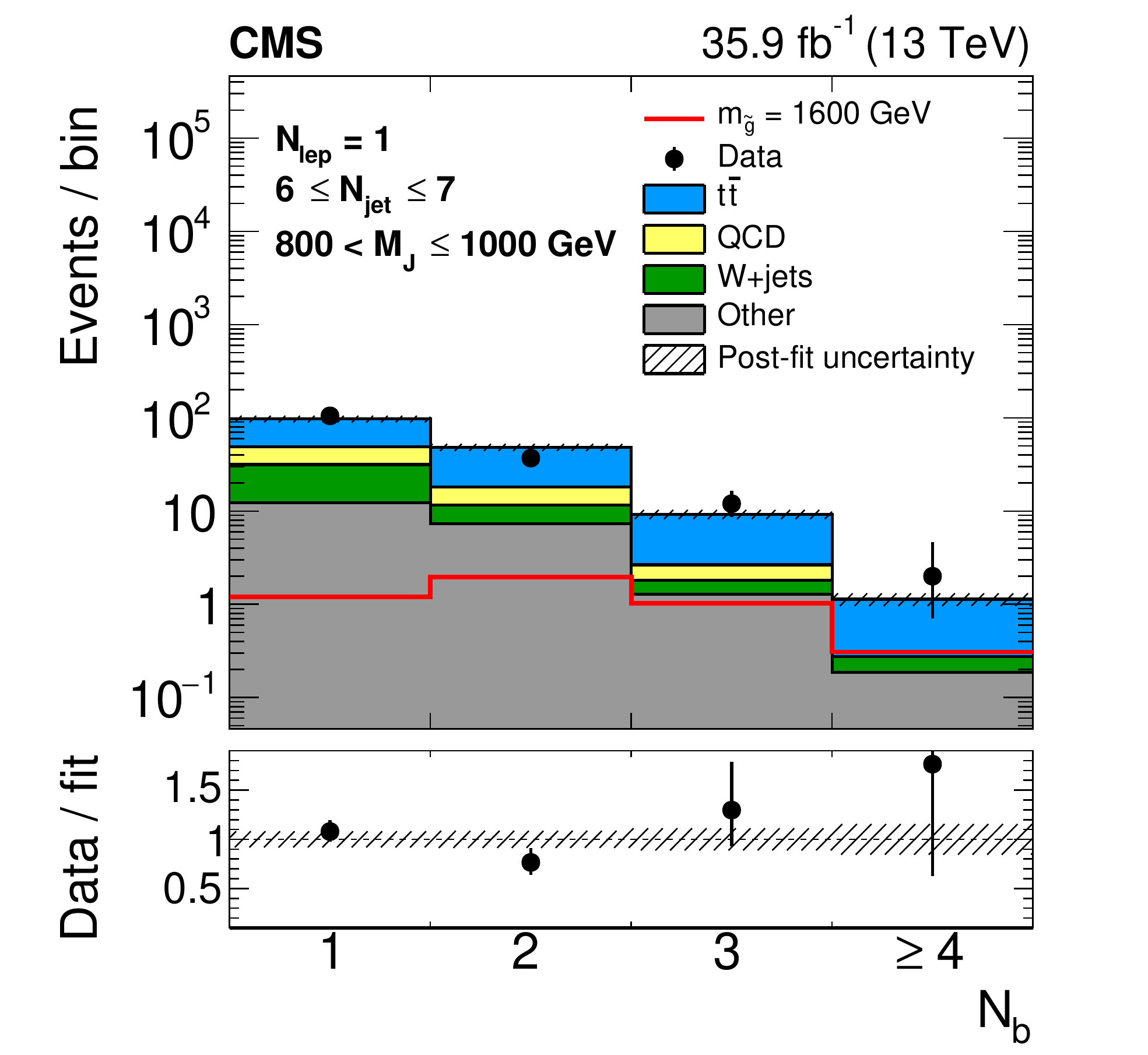} \hfil
\includegraphics[width=\cmsFigWidth]{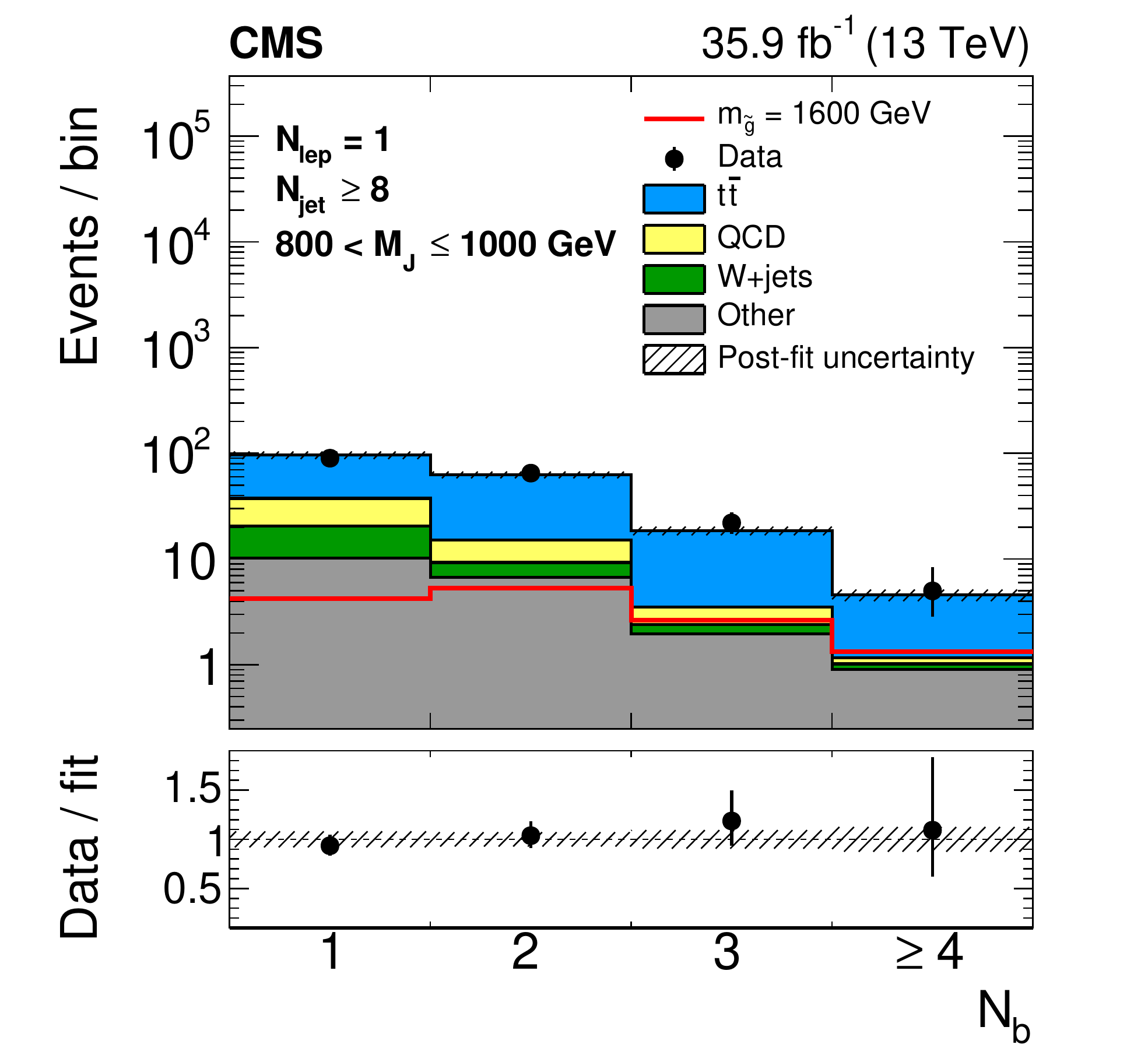} \\
\includegraphics[width=\cmsFigWidth]{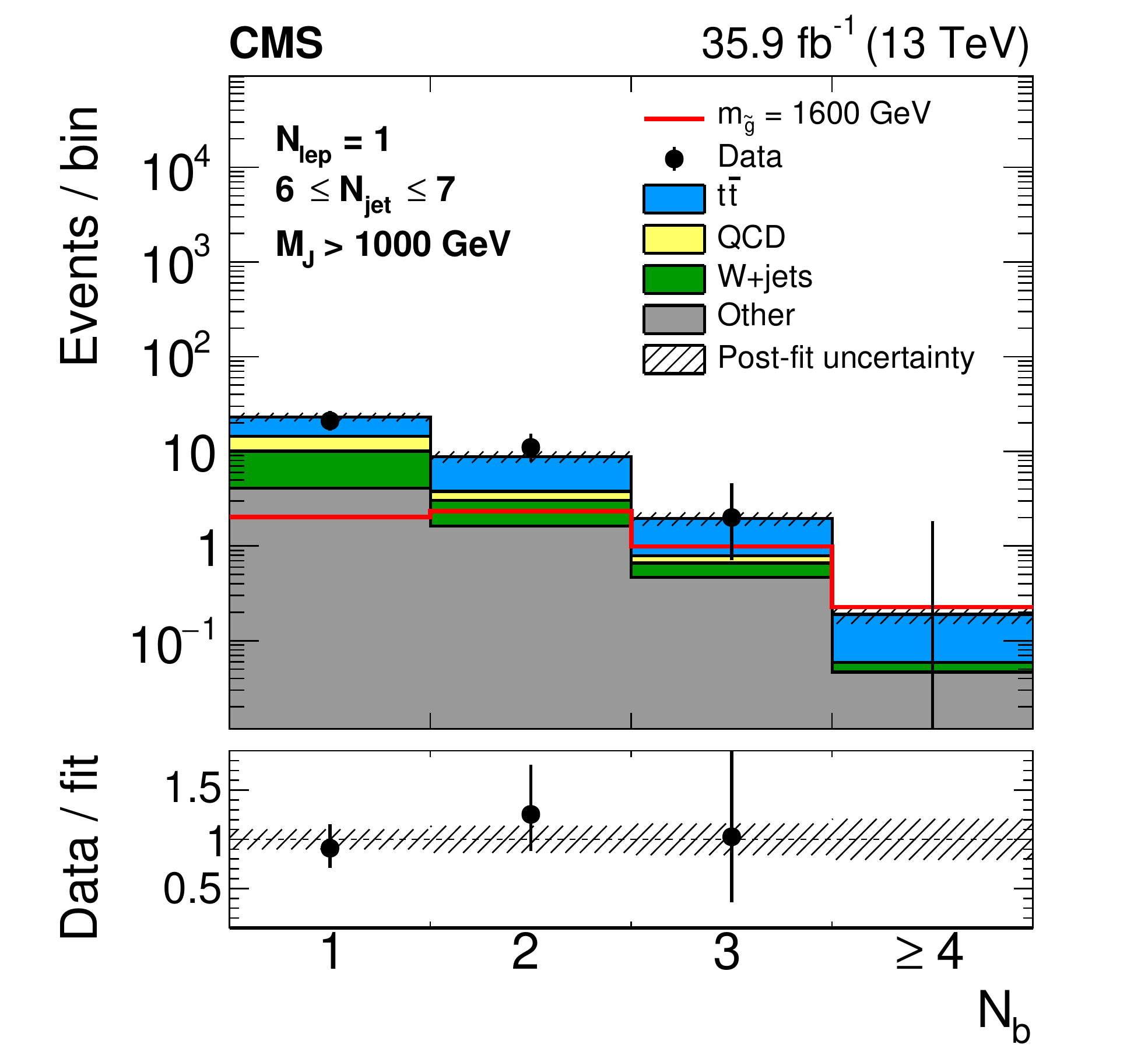} \hfil
\includegraphics[width=\cmsFigWidth]{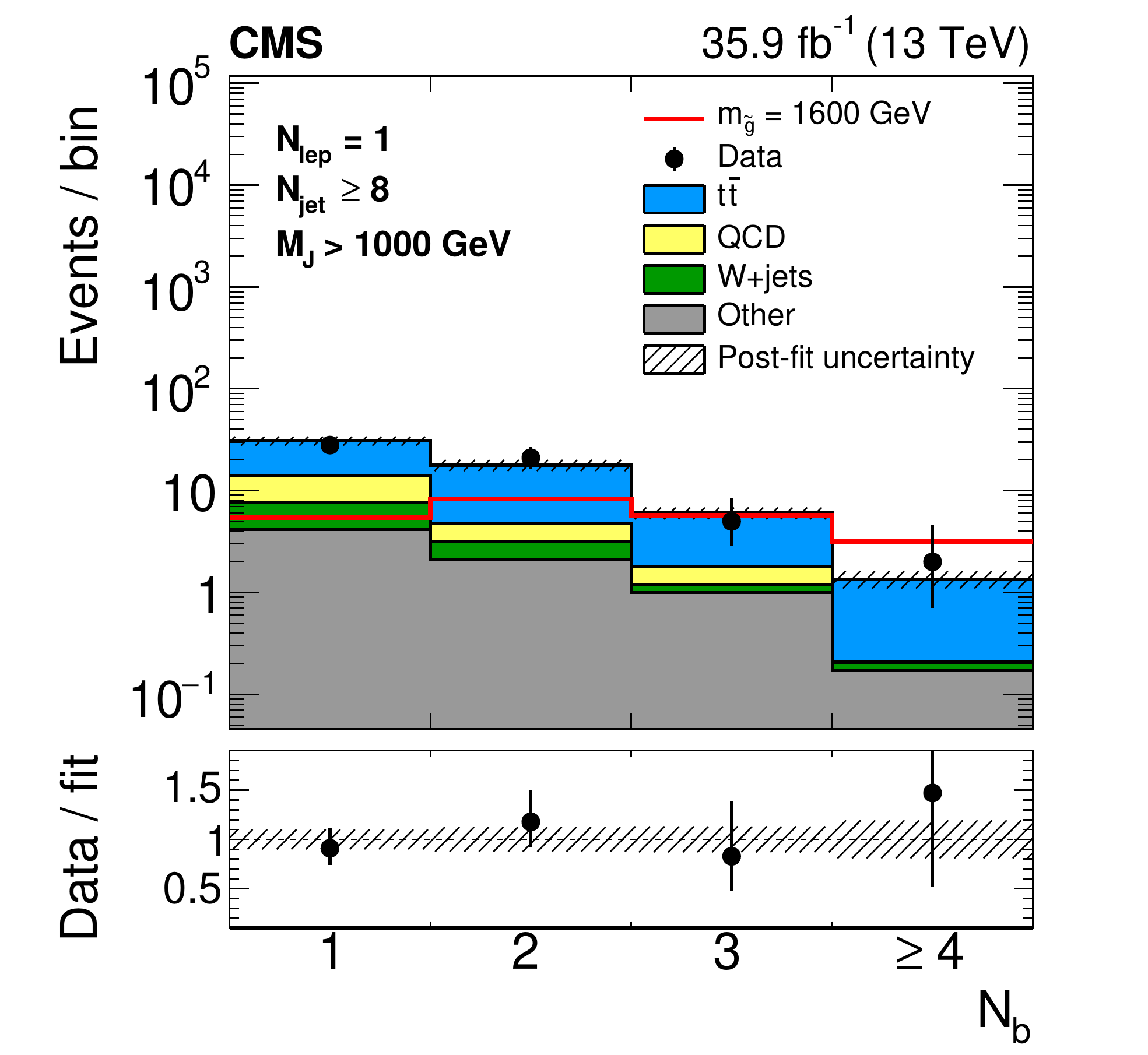}

\caption{Data and the background-only post-fit \Nb distribution for bins with large expected signal contribution: $800 < \MJ \leq 1000\GeV$, $ 6 \leq \Njet \leq 7$ (upper-left), $800 < \MJ \leq 1000\GeV$, $\Njet \geq 8$ (upper-right), $\MJ > 1000\GeV$, $ 6 \leq \Njet \leq 7$ (lower-left), and $\MJ > 1000\GeV$, $ \Njet \geq 8$ (lower-right).
The expected signal distribution is also shown for a gluino mass of 1600\GeV.
The ratio of data to post-fit yields is shown in the lower panel.
The post-fit uncertainty is depicted as a hatched band.}
\label{fig:bonly_sr}
\end{figure*}

The results of a background-only fit of the observed \Nb distributions are shown in
Figs.~\ref{fig:bonly_cr} and \ref{fig:bonly_sr}.
These figures separately show the $\Nlep=1$ control and signal regions,
although the fit includes all bins simultaneously.
The \Nb distributions in data are well described by the fit,
and examination of the nuisance parameters shows that none
of them are significantly changed by the fit.
The post-fit yields are presented in Table~\ref{tab:bonly_yields}.

\begin{table*}
\centering
\topcaption{\label{tab:bonly_yields} Post-fit yields for the background-only fit, observed data, and expected yields for $m_{\PSg}=1600\GeV$ in each search bin.}
\begin{tabular}[tbp!]{ l | c  c  c  c | c |  c | c  }
\hline
$\Nb$ & QCD & $\cPqt\cPaqt$ & \PW+jets & Other & All bkg. & Data & Expected $m_{\PSg}=1600\GeV$\\
\hline
\multicolumn{8}{c}{$4\leq \Njet\leq5,~500<\MJ\leq800\GeV$}\\
\hline
$1$ & $148$ & $340$ & $196$ & $91$ & $775\pm43\x$ & $777$ & $0.50 \pm 0.13$ \\
$2$ & $29$ & $175$ & $30$ & $31$ & $264\pm17\x$ & $264$ & $0.39 \pm 0.11$ \\
$3$ & $4.3$ & $24.8$ & $2.5$ & $4.4$ & $36\pm4\x$ & $34$ &  $0.18 \pm 0.08$ \\
${\geq} 4$ & $0.0$ & $2.2$ & $0.3$ & $0.2$ & $2.7\pm0.4$ & $3$ & $0.04 \pm 0.04$ \\
\hline
\multicolumn{8}{c}{$4\leq \Njet\leq5,~\MJ>800\GeV$}\\
\hline
$1$ & $16.5$ & $26.3$ & $22.5$ & $11.0$ & $76\pm6\x$ & $77$ & $0.32 \pm 0.11$ \\
$2$ & $1.1$ & $10.6$ & $3.4$ & $3.8$ & $19\pm2\x$ & $18$ & $0.40 \pm 0.12$ \\
$3$ & $0.7$ & $1.3$ & $0.3$ & $0.3$ & $2.7\pm0.5$ & $3$ & $0.13 \pm 0.06$ \\
${\geq} 4$ & $0.00$ & $0.09$ & $0.03$ & $0.01$ & $0.13\pm0.03$ & $0$ & $0.03 \pm 0.03$ \\
\hline
\multicolumn{8}{c}{$6\leq \Njet\leq7,~500<\MJ\leq800\GeV$}\\
\hline
$1$ & $197$ & $620$ & $169$ & $120$ & $1106\pm48\x\x$ & $1105$ &  $2.5 \pm 0.3$ \\
$2$ & $49$ & $440$ & $36$ & $66$ & $591\pm21\x$ & $588$ & $3.1 \pm 0.3$ \\
$3$ & $6.4$ & $89.2$ & $4.6$ & $13.4$ & $114\pm8\x\x$ & $112$ & $1.4 \pm 0.2$ \\
${\geq} 4$ & $1.9$ & $11.4$ & $0.6$ & $2.1$ & $16\pm2\x$ & $21$ & $0.25 \pm 0.09$ \\
\hline
\multicolumn{8}{c}{$\Njet\geq8,~500<\MJ\leq800\GeV$}\\
\hline
$1$ & $130$ & $574$ & $53$ & $68$ & $825\pm38\x$ & $821$ & $3.5 \pm 0.3$ \\
$2$ & $45$ & $478$ & $14$ & $49$ & $586\pm20\x$ & $603$ & $5.4 \pm 0.4$ \\
$3$ & $6.3$ & $138.1$ & $2.5$ & $16.7$ & $164\pm9\x\x$ & $148$ & $3.0 \pm 0.3$ \\
${\geq} 4$ & $2.8$ & $29.8$ & $0.4$ & $4.8$ & $38\pm4\x$ & $40$ &  $1.4 \pm 0.2$ \\
\hline
\multicolumn{8}{c}{$6\leq \Njet\leq7,~800<\MJ\leq1000\GeV$}\\
\hline
$1$ & $17.3$ & $48.4$ & $19.2$ & $12.3$ & $97\pm8\x$ & $105$ & $1.2 \pm 0.2$ \\
$2$ & $6.6$ & $30.1$ & $4.3$ & $7.3$ & $48\pm4\x$ & $37$ & $2.0 \pm 0.3$ \\
$3$ & $0.8$ & $6.6$ & $0.5$ & $1.3$ & $9.3\pm1.0$ & $12$ & $1.0 \pm 0.2$ \\
${\geq} 4$ & $0.0$ & $0.9$ & $0.1$ & $0.2$ & $1.1\pm0.2$ & $2$ & $0.31 \pm 0.09$ \\
\hline
\multicolumn{8}{c}{$\Njet\geq8,~800<\MJ\leq1000\GeV$}\\
\hline
$1$ & $17.0$ & $58.7$ & $10.3$ & $10.2$ & $96\pm8\x$ & $90$ & $4.2 \pm 0.4$ \\
$2$ & $5.8$ & $47.5$ & $2.5$ & $6.8$ & $63\pm5\x$ & $65$ & $5.3 \pm 0.4$ \\
$3$ & $1.1$ & $15.0$ & $0.4$ & $2.0$ & $19\pm2\x$ & $22$ & $2.6 \pm 0.3$ \\
${\geq} 4$ & $0.2$ & $3.4$ & $0.1$ & $0.9$ & $4.6\pm0.6$ & $5$ & $1.3 \pm 0.2$ \\
\hline
\multicolumn{8}{c}{$6\leq \Njet\leq7,~\MJ>1000\GeV$}\\
\hline
$1$ & $4.4$ & $8.7$ & $6.0$ & $4.1$ & $23\pm2\x$ & $21$ & $2.0 \pm 0.3$ \\
$2$ & $0.7$ & $5.0$ & $1.4$ & $1.6$ & $8.8\pm1.2$ & $11$ & $2.3 \pm 0.3$ \\
$3$ & $0.1$ & $1.2$ & $0.2$ & $0.5$ & $1.9\pm0.3$ & $2$ & $1.0 \pm 0.2$ \\
${\geq} 4$ & $0.00$ & $0.13$ & $0.01$ & $0.05$ & $0.19\pm0.04$ & $0$ & $0.23 \pm 0.08$ \\
\hline
\multicolumn{8}{c}{$\Njet\geq8,~\MJ>1000\GeV$}\\
\hline
$1$ & $6.4$ & $16.7$ & $3.5$ & $4.1$ & $31\pm3\x$ & $28$ &  $5.4 \pm 0.4$ \\
$2$ & $1.6$ & $13.1$ & $1.1$ & $2.1$ & $18\pm2\x$ & $21$ & $8.2 \pm 0.5$ \\
$3$ & $0.6$ & $4.2$ & $0.2$ & $1.0$ & $6.0\pm0.8$ & $5$ & $5.7 \pm 0.4$ \\
${\geq} 4$ & $0.0$ & $1.2$ & $0.0$ & $0.2$ & $1.4\pm0.3$ & $2$ & $3.2 \pm 0.3$  \\
\hline
\end{tabular}
\end{table*}

\begin{figure*}[tbp!]
\centering
\includegraphics[width=\cmsFigWidthLarge]{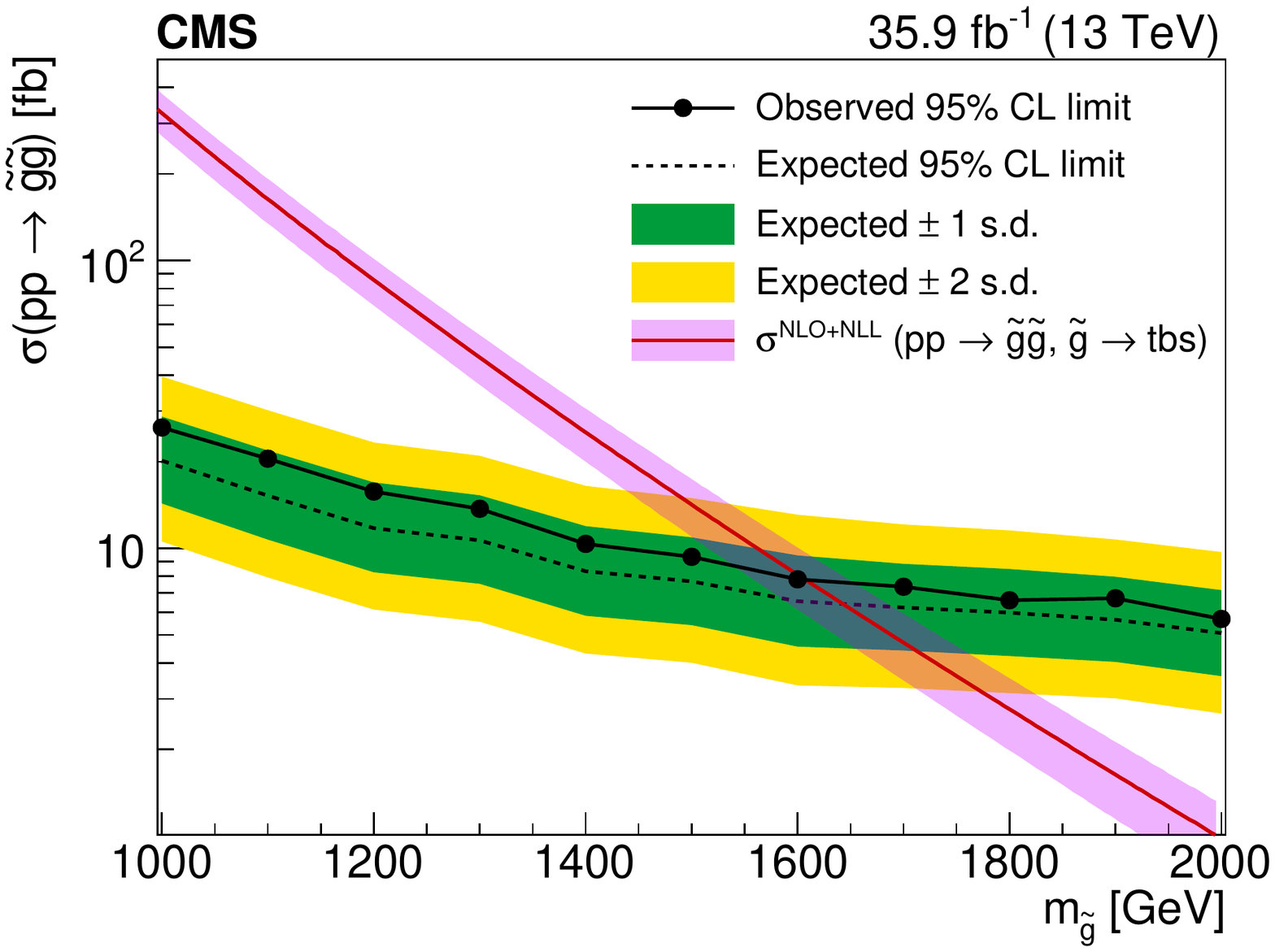}
\caption{Cross section upper limits at 95\% CL for a model of gluino pair production with $\PSg\to\cPqt\cPqb\cPqs$ compared to the gluino pair production cross section.
The theoretical uncertainties in the cross section are shown as a band around the
red line~\cite{XSecgluinogluino}.
The expected limits (dashed line) and their ${\pm}1$~s.d.\ and ${\pm}2$~s.d.\ variations are shown as green and yellow bands, respectively.
The observed limit is shown by the solid line with dots.}
\label{fig:limit}
\end{figure*}

A signal-plus-background fit is performed for gluino masses ranging from 1000 to 2000\GeV.
For all masses, the post-fit \Nb distribution describes the data well, and the fit extracts at most a small and insignificant signal contribution. For example, with a 1600\GeV gluino, the extracted signal yield relative to the model prediction is $r=0.18^{+0.41}_{-0.18}$.
The change of nuisance parameters by the fit is small and consistent with those of the background-only fit.
Limits on the signal production cross section are calculated at 95\% confidence level (CL) using the asymptotic approximation of the $\textrm{CL}_\textrm{s}$ criterion~\cite{0954-3899-28-10-313,CMS-NOTE-2011-005,Cowan:2010js,Junk:1999kv} and shown in Fig.~\ref{fig:limit}.
Comparing the observed limit to the gluino pair production cross section~\cite{XSecgluinogluino}, gluino masses below 1610\GeV are excluded in the benchmark $\PSg \to \cPqt\cPqb\cPqs$ model.

\section{Summary}
\label{sec:Summary}

Results are presented from a search for new phenomena in events with a single lepton, large jet and bottom quark jet multiplicities, and high sum of large-radius jet masses, without a missing transverse momentum requirement.
The background is predicted using a simultaneous fit in bins of the number of jets, number of \cPqb-tagged jets, and the sum of masses of large radius jets, using
Monte Carlo simulated predictions with corrections measured in data control samples for the normalizations of the dominant backgrounds and nuisance parameters for theoretical and experimental uncertainties.
Statistical uncertainties dominate in the signal regions,
while the most important systematic uncertainties arise from the modeling of gluon splitting
and the \cPqb quark tagging efficiency and mistag rate.
The observed data are consistent with the background-only hypothesis.
An upper limit of approximately 10~fb is determined for the gluino-gluino production cross section using a benchmark $R$-parity violating supersymmetry model of gluino pair production with a prompt three-body decay to $\cPqt\cPqb\cPqs$ quarks, as predicted in minimal flavor violating models.
For this model, gluinos are observed (expected) to be excluded up to $1610\ (1640)\GeV$ at a 95\% confidence level, which improves upon previous searches at $\sqrt{s}=8\TeV$~\cite{Aad:2016kww,Khachatryan:2016iqn,Khachatryan:2016unx} and is comparable to recent results at $13\TeV$~\cite{1leprpv_atlas_2016}.

\begin{acknowledgments}

We congratulate our colleagues in the CERN accelerator departments for the excellent performance of the LHC and thank the technical and administrative staffs at CERN and at other CMS institutes for their contributions to the success of the CMS effort. In addition, we gratefully acknowledge the computing centers and personnel of the Worldwide LHC Computing Grid for delivering so effectively the computing infrastructure essential to our analyses. Finally, we acknowledge the enduring support for the construction and operation of the LHC and the CMS detector provided by the following funding agencies: BMWFW and FWF (Austria); FNRS and FWO (Belgium); CNPq, CAPES, FAPERJ, and FAPESP (Brazil); MES (Bulgaria); CERN; CAS, MoST, and NSFC (China); COLCIENCIAS (Colombia); MSES and CSF (Croatia); RPF (Cyprus); SENESCYT (Ecuador); MoER, ERC IUT, and ERDF (Estonia); Academy of Finland, MEC, and HIP (Finland); CEA and CNRS/IN2P3 (France); BMBF, DFG, and HGF (Germany); GSRT (Greece); OTKA and NIH (Hungary); DAE and DST (India); IPM (Iran); SFI (Ireland); INFN (Italy); MSIP and NRF (Republic of Korea); LAS (Lithuania); MOE and UM (Malaysia); BUAP, CINVESTAV, CONACYT, LNS, SEP, and UASLP-FAI (Mexico); MBIE (New Zealand); PAEC (Pakistan); MSHE and NSC (Poland); FCT (Portugal); JINR (Dubna); MON, RosAtom, RAS, RFBR and RAEP (Russia); MESTD (Serbia); SEIDI, CPAN, PCTI and FEDER (Spain); Swiss Funding Agencies (Switzerland); MST (Taipei); ThEPCenter, IPST, STAR, and NSTDA (Thailand); TUBITAK and TAEK (Turkey); NASU and SFFR (Ukraine); STFC (United Kingdom); DOE and NSF (USA).

\hyphenation{Rachada-pisek} Individuals have received support from the Marie-Curie program and the European Research Council and Horizon 2020 Grant, contract No. 675440 (European Union); the Leventis Foundation; the A. P. Sloan Foundation; the Alexander von Humboldt Foundation; the Belgian Federal Science Policy Office; the Fonds pour la Formation \`a la Recherche dans l'Industrie et dans l'Agriculture (FRIA-Belgium); the Agentschap voor Innovatie door Wetenschap en Technologie (IWT-Belgium); the Ministry of Education, Youth and Sports (MEYS) of the Czech Republic; the Council of Science and Industrial Research, India; the HOMING PLUS program of the Foundation for Polish Science, cofinanced from European Union, Regional Development Fund, the Mobility Plus program of the Ministry of Science and Higher Education, the National Science Center (Poland), contracts Harmonia 2014/14/M/ST2/00428, Opus 2014/13/B/ST2/02543, 2014/15/B/ST2/03998, and 2015/19/B/ST2/02861, Sonata-bis 2012/07/E/ST2/01406; the National Priorities Research Program by Qatar National Research Fund; the Programa Clar\'in-COFUND del Principado de Asturias; the Thalis and Aristeia programs cofinanced by EU-ESF and the Greek NSRF; the Rachadapisek Sompot Fund for Postdoctoral Fellowship, Chulalongkorn University and the Chulalongkorn Academic into Its 2nd Century Project Advancement Project (Thailand); the Welch Foundation, contract C-1845; and the Weston Havens Foundation (USA).
\end{acknowledgments}

\ifthenelse{\boolean{cms@external}}{}{\clearpage}
\bibliography{auto_generated}

\cleardoublepage \appendix\section{The CMS Collaboration \label{app:collab}}\begin{sloppypar}\hyphenpenalty=5000\widowpenalty=500\clubpenalty=5000\vskip\cmsinstskip
\textbf{Yerevan Physics Institute,  Yerevan,  Armenia}\\*[0pt]
A.M.~Sirunyan,  A.~Tumasyan
\vskip\cmsinstskip
\textbf{Institut f\"{u}r Hochenergiephysik,  Wien,  Austria}\\*[0pt]
W.~Adam,  F.~Ambrogi,  E.~Asilar,  T.~Bergauer,  J.~Brandstetter,  E.~Brondolin,  M.~Dragicevic,  J.~Er\"{o},  A.~Escalante Del Valle,  M.~Flechl,  M.~Friedl,  R.~Fr\"{u}hwirth\cmsAuthorMark{1},  V.M.~Ghete,  J.~Grossmann,  J.~Hrubec,  M.~Jeitler\cmsAuthorMark{1},  A.~K\"{o}nig,  N.~Krammer,  I.~Kr\"{a}tschmer,  D.~Liko,  T.~Madlener,  I.~Mikulec,  E.~Pree,  N.~Rad,  H.~Rohringer,  J.~Schieck\cmsAuthorMark{1},  R.~Sch\"{o}fbeck,  M.~Spanring,  D.~Spitzbart,  A.~Taurok,  W.~Waltenberger,  J.~Wittmann,  C.-E.~Wulz\cmsAuthorMark{1},  M.~Zarucki
% ...
\vskip\cmsinstskip
% ...
1:~Also at Vienna University of Technology,  Vienna,  Austria\\
% ...
\end{sloppypar}
\end{document}